\documentclass{birkmult}
\newtheorem{thm}{Theorem}[section]

\newtheorem{lemma}[thm]{Lemma}
\newtheorem{prop}[thm]{Proposition}
\theoremstyle{definition}
\newtheorem{defn}[thm]{Definition}
\theoremstyle{remark}

\numberwithin{equation}{section}
\usepackage{amsmath}
\usepackage{amssymb}
\usepackage{amsthm}
\usepackage{diagxy}
\usepackage[all]{xy}
\usepackage{epsfig}
\usepackage{rotating}
\newcommand{\Super}{\mathcal{S}}
\newcommand{\SuperRes}{\mathcal{S}_{\rm F}}
\newcommand{\MCG}{\mathcal{G}(\Sigma)}
\newcommand{\MCGInf}{\mathcal{G}_{\infty}(\Sigma)}
\newcommand{\MCGF}{\mathcal{G}_{\rm F}(\Sigma)}
\newcommand{\sss}{\scriptscriptstyle}
\newcommand{\Diff}{\text{Diff}(\Sigma)}
\newcommand{\DiffId}{\text{Diff}^0(\Sigma)}
\newcommand{\DiffInf}{\text{Diff}_\infty(\Sigma)}
\newcommand{\DiffInfId}{\text{Diff}^0_\infty(\Sigma)}
\newcommand{\DiffF}{\text{Diff}_{\rm F}(\Sigma)}
\newcommand{\DiffFId}{\text{Diff}^0_{\rm F}(\Sigma)}
\newcommand{\Riem}{\text{Riem}(\Sigma)}
\newcommand{\RP}{\mathbb{R}\mathrm{P}}
\newcommand{\image}{\mathrm{image}}
\newcommand{\kernel}{\mathrm{kernel}}
\renewcommand{\square}{\Box}

\newcommand{\leftnormal}{\lhd}

\newcommand{\rightnormal}{\rhd}

\begin{document}
\title[Mapping-class groups]%
{Mapping-class groups of 3-manifolds in\\ 
 canonical quantum gravity}

\author[D. Giulini]{Domenico Giulini}

\address{%
Physikalisches Institut     \\
Universit\"at Freiburg      \\
Hermann-Herder-Stra{\ss}e 3 \\
D-79104 Freiburg            \\
Germany}
\email{giulini@physik.uni-freiburg.de}

\thanks{Contribution to the proceedings of the workshop on 
\emph{mathematical and physical aspects of quantum gravity} held in 
Blaubeuren/Germany from July\,28th to August\,1st 2005. I sincerely thank Bertfried 
Fauser and J\"urgen Tolksdorf for organizing Blaubeuren\,II and 
inviting me to it.
}



\date{February 15, 2006}

\dedicatory{This contribution is dedicated to Rafael Sorkin 
on the occasion of his 60th birthday}

\begin{abstract}
Mapping-class groups of 3-manifolds feature as symmetry groups 
in canonical quantum gravity. They are an obvious source through 
which topological information could be transmitted into the quantum 
theory. If treated as gauge symmetries, their inequivalent 
unitary irreducible representations should give rise to a complex 
superselection structure. This highlights certain aspects of 
spatial diffeomorphism invariance that to some degree seem physically 
meaningful and which persist in all approaches based on smooth 
3-manifolds, like geometrodynamics and loop quantum gravity. 
We also attempt to give a flavor of the mathematical ideas 
involved. 
\end{abstract}

\maketitle

\section{Some facts about Hamiltonian General Relativity}
\label{sec:FactsHGR}
\subsection{Introduction}
\label{sec:Introduction}
As is well known, Einstein's field equations for General Relativity 
can be cast into the form of a constrained Hamiltonian system. The 
unreduced phase space is then given by the cotangent bundle over 
the space $\Riem$ of Riemannian metrics on a 3-manifold $\Sigma$.
This phase space is coordinatized by $(q,p)$, where $q$ is a 
Riemannian metric on $\Sigma$ and $p$ is a section in the bundle 
of symmetric contravariant tensors of rank two and density-weight 
one over $\Sigma$. 

The relation of these objects to the description of a solution to 
Einstein's equations in terms of a four-dimensional globally 
hyperbolic Lorentzian manifold $(g,M)$ is as follows: let $\Sigma$ be 
a Cauchy surface in $M$ and $q:=g\vert_{\sss T(M)}$ its induced 
metric (first fundamental form). Let $n$ be the normal field to 
$\Sigma$, where $g(n,n)=-1$, i.e. we use the `mostly plus' 
signature convention where timelike vectors have a negative 
$g$-square. Let $D$ and $\nabla$ be the Levi-Civita covariant 
derivatives on $(q,\Sigma)$ and $(g,M)$ respectively. Then for any 
vector $X$ tangent to $\Sigma$ and any vector field $\tilde Y$ on 
$M$ whose restriction $Y$ to $\Sigma$ is tangential to $\Sigma$, we have   
\begin{equation}
\label{eq:DefExtCurv}
\nabla_X\tilde Y=D_X Y+n\,K(X,Y)\,,
\end{equation}
were $n\,K(X,Y)$ represents the normal component of $\nabla_X\tilde Y$. 
It is easy to see that $K$ is a symmetric covariant tensor of rank\,2 
on $\Sigma$, also called the extrinsic curvature (or second fundamental 
form) of $\Sigma$ in $M$. The canonical momentum field 
$p$ can now be expressed in terms of these data: 
\begin{equation}
\label{eq:MomentumExtCurv}
p=
\sqrt{\det(q)}\,\bigl(K-q\,\text{Tr}_q(K)\bigr)^\sharp\,.
\end{equation}
Here $\sharp$ denotes the isomorphism $T^*(M)\rightarrow T(M)$ 
induced by $q$ (`index raising'), extended to tensors of all ranks. 

Einstein's equations in Hamiltonian form now decompose into two 
evolution equations (of six independent component equations each): 
\begin{subequations}
\label{eq:EvolEq}
\begin{alignat}{2}
\label{eq:EvolEq1}
& \dot q &&\,=\, F_1(q,p;\alpha,\beta;\phi)\,,\\
\label{eq:EvolEq2}
& \dot p &&\,=\, F_2(q,p;\alpha,\beta;\phi)\,,
\end{alignat}
\end{subequations}
and two equations without time derivatives in $q$ and $p$ 
(of one and three independent component equations respectively),
thereby implying constraints on the initial data: 
\begin{subequations}
\label{eq:ConstEq}
\begin{alignat}{2}
\label{eq:ConstEq1}
&C_s(q,p;\phi)&&\,
:=\,G_q(p,p)-\sqrt{\det(q)}\,\bigl(S(q)-2\Lambda\bigr)
  +2\rho_m(\phi,q)\,=\,0\,,\\
\label{eq:ConstEq2}
&C_v(q,p;\phi)&&\,
:=\,-2\text{div}_qp+j_m(\phi,q)\,=\,0\,.
\end{alignat}
\end{subequations}
These are referred to as the scalar and the vector (or diffeomorphism) 
constraints respectively.
  
The meanings of the symbols in (\ref{eq:EvolEq}) and (\ref{eq:ConstEq}) 
are as follows: $F_{1,2}$ are local functionals whose explicit forms 
need not interest us at this moment. $\alpha$ and $\beta$ are a scalar 
function and a vector field on $\Sigma$ respectively (the `lapse' and 
`shift' function) which are not determined by the equations of motion 
but which one needs to specify by hand. They represent the four free 
functions out of the ten component functions $g_{\mu\nu}$ which are 
\emph{not} 
determined by the equations of motion due to spacetime-diffeomorphism 
invariance. $S(q)$ is the scalar curvature (Ricci scalar) of 
$(q,\Sigma)$ and $\text{div}_q$ denotes the covariant divergence 
with respect to the Levi Civita derivative on $(q,\Sigma)$, 
i.e. in components $(\text{div}_qp)^b=D_ap^{ab}$. The symbol $\phi$ 
collectively represents possible matter fields. $\rho_m$ and $j_m$ 
are respectively the scalar and vector densities of weight one 
for the energy and momentum density of the matter. As usual, 
$\Lambda$ is the cosmological constant. Finally, $G_q$ is a bilinear
form, the so-called DeWitt metric, that maps a pair of symmetric
contravariant 2nd-rank tensors of density weight one to a 
scalar density of weight one. In components one has 
\begin{equation}
\label{eq:DeWittMetric}
\begin{split}
G_q(p_1,p_2)
&=\bigl[\det(q_{ab})\bigr]^{-1/2}\,\tfrac{1}{2}
\bigl(q_{ac}q_{bd}+q_{ad}q_{bc}-q_{ab}q_{cd}\bigr)p_1^{ab}p_2^{cd}\\
&=\bigl[\det(q_{ab})\bigr]^{-1/2}\bigl(p_1^{ab}p_{2\,ab}-\tfrac{1}{2}
p^a_{1\,a}p^b_{2\,b}\bigr)\,.\\
\end{split}
\end{equation}
Pointwise (on $\Sigma$) it defines a Lorentzian metric of signature
$(1,5)$ (the `negative direction' being the trace mode) on the six 
dimensional space of symmetric 2nd-rank tensor densities, a discussion 
of which may be found in~\cite{DeWittQTGI:1967}. Some relevant aspects 
of the infinite-dimensional geometry that is obtained by integrating 
$G_q(p_1,p_2)$ over $\Sigma$ (the so-called `Wheeler-DeWitt metric') 
are discussed in~\cite{Giulini:1995c}.

\subsection{Topologically closed Cauchy surfaces}
\label{sec:TopClosed}
If $\Sigma$ is closed (compact without boundary) the constraints
(\ref{eq:ConstEq}) actually generate all of the dynamical evolution
(\ref{eq:EvolEq}). That is, we can write 
\begin{subequations}
\label{eq:EvolEqHam}
\begin{alignat}{3}
\label{eq:EvolEqHam1}
&F_1&&\,=\,&&\frac{\delta H}{\delta p}\,,\\
\label{eq:EvolEqHam2}
&F_2&&\,=\,-\,&&\frac{\delta H}{\delta q}\,,
\end{alignat}
\end{subequations}
where
\begin{equation}
\label{eq:TotalHamiltonianClosed}
H[q,p;\alpha,\beta;\phi]
=\int_\Sigma\alpha C_s(q,p;\phi)
+\int_\Sigma\beta\cdot C_v(q,p;\phi)\,.
\end{equation}
Here $\beta\cdot C_v$ denote the natural pairing between a vector
($\beta$) and a one-form of density weight one ($C_v$).  
This means that the entire dynamical evolution is generated by 
constraints. These constraints form a first-class system, which 
means that the Poisson bracket of two of them is again a linear 
combination (generally with phase-space dependent coefficients) 
of the  constraints. Writing 
\begin{equation}
\label{eq:SmearedConstraints}
C_s(\alpha):=\int_\Sigma \alpha C_s\,,\quad\text{and}\quad
C_v(\beta):=\int_\Sigma \beta\cdot C_v
\end{equation}
we have 
\begin{subequations}
\label{eq:Constraints}
\begin{alignat}{2}
\label{eq:Constraints1}
& \big\{C_s(\alpha)\,,\,C_s(\alpha')\big\}
&&\,=\,C_v\bigl(\alpha (d\alpha')^\sharp-\alpha'(d\alpha)^\sharp\bigr)\,,\\
\label{eq:Constraints2}
& \bigl\{C_v(\beta)\,,\,C_s(\alpha)\bigr\}
&&\,=\,C_s\bigl(\beta\cdot d\alpha\bigr)\,,\\
\label{eq:Constraints3}
& \bigl\{C_v(\beta)\,,\,C_v(\beta')\bigr\}
&&\,=\,C_v\bigl([\beta,\beta']\bigr) \,.
\end{alignat}
\end{subequations}
In passing we remark that (\ref{eq:Constraints3}) says that the vector 
constraints form a Lie-subalgebra which, however, according to 
(\ref{eq:Constraints2}), is \emph{not} an ideal. This means that the 
flows generated by the scalar constraints are not tangential to 
the constraint-hypersurface that is determined by the vanishing of 
the vector constraints, except for the points where the 
constraint-hypersurfaces for the scalar and vector constraints 
intersect. This means that generally one cannot reduce the constraints 
in steps: \emph{first}  the vector constraints and \emph{then} the 
scalar constraints, simply because the scalar constraints do not 
act on the solution space of the vector constraints. This difficulty 
clearly persists in any implementation of (\ref{eq:Constraints}) 
as operator constraints in canonical quantum gravity.  

According to the orthodox theory of constrained 
systems~\cite{Dirac:LQM,HenneauxTeitelboim:QGS}, all motions 
generated by the \emph{first class} constraints should be interpreted 
as gauge transformations, i.e. be physically unobservable.\footnote{%
In \cite{Dirac:LQM} Dirac proposed this in the form of a 
conjecture. It has become the orthodox view that is also 
adopted in \cite{HenneauxTeitelboim:QGS}. There are simple and well 
known---though rather pathological---counterexamples (e.g.  
\cite{HenneauxTeitelboim:QGS} \S\,1.2.2 and \S\,1.6.3). 
The conjecture can be proven under the hypothesis that there are 
no \emph{ineffective} constraints, i.e. whose Hamiltonian vector fields 
vanish on the constraint hypersurface (e.g. 
\cite{CaboLouis-Martinez:1990}; also \cite{HenneauxTeitelboim:QGS} 
\S\,3.3.2). For further discussion of this rather subtle point 
see \cite{Gotay:1983} and \cite{GotayNester:1984}. Note however 
that these issues  are only concerned with the algebraic form in which 
the constraints are delivered by the formalism. For example, if 
$\phi(q,p)=0$ is effective (i.e $d\phi\vert_{\phi^{-1}(0)}\ne 0$) 
then $\phi^2(q,p)=0$ is clearly ineffective, even though it defines the 
\emph{same} constraint subset in phase space. In a proper geometric 
formulation the algebra of observables just depends on this constraint
subset: define the `gauge Poisson-algebra', Gau,  by the set of all 
smooth functions that vanish on this set (it is clearly an ideal 
with respect to pointwise multiplication, but not a Lie-ideal). Then 
take as algebra of physical observables the quotient of the Lie 
idealizer of Gau (in the Poisson algebra of, say, smooth functions 
on unconstrained phase space) with respect to Gau. See e.g. 
\cite{Giulini:2003} for more details.} In other words, 
states connected by a motion that is generated by first-class 
constraints are to be considered as \emph{physically identical}. 

The conceptual question of how one should interpret the fact that 
all evolution is pure gauge is know as the \emph{problem of time} in 
classical and also in quantum general relativity. It is basically 
connected with the constraints $C_s(\alpha)$, since their 
Hamiltonian flow represents a change on the canonical variables 
that corresponds to the motion of the hypersurface $\Sigma$ in $M$ 
in normal direction. For a detailed discussion see Section\,5.2 in 
\cite{Kiefer:QuantumGravity}. 

In contrast, the meaning of the flow generated by the 
constraints $C_v(\beta)$ is easy to understand: it just corresponds 
to an infinitesimal diffeomorphism within $\Sigma$. Accordingly, its 
action on a local\footnote{`Local' meaning that the real-valued 
function $F$ on $\Sigma$ depends on $x\in\Sigma$ through the 
values of $q$ and $p$ as well their derivatives up to \emph{finite} 
order at $x$.} phase-space function $F[q,p](x)$ is just given by 
its Lie derivative: 
\begin{equation}
\label{eq:DiffConstAction}
\big\{F,C_v(\beta)\big\}=L_\beta F\,.
\end{equation}
Hence the gauge group generated by the vector constraints is the 
identity component $\DiffId$ of the diffeomorphism group $\Diff$
of $\Sigma$. Note that this is true despite the fact that 
$\Diff$ is only a Fr\'echet Lie group and that, accordingly, the 
exponential map is not surjective on any neighborhood of the 
identity (cf.~\cite{Freifeld:1968}). The point being that  
$\DiffId$ is simple (cf. \cite{McDuff:1978}) and that the 
subgroup generated\footnote{The subgroup `generated' by a set 
is the subgroup of all finite products of elements in this set.} 
by the image of the exponential map is clearly a non-trivial
normal subgroup of, and hence equivalent to, $\DiffId$.
     
What about those transformations in $\Diff$ which are not in the 
identity component (i.e. the so called `large' diffeomorphisms)?
Are they, too, to be looked at as pure gauge transformations, or 
are they physically meaningful (observable) symmetries? Suppose 
we succeeded in constructing the reduced phase space with respect 
to $\DiffId$, we would then still have a residual non-trivial 
action of the discrete group
\begin{equation}
\label{eq:DefMCG}
\MCG:=\Diff/\DiffId=:\pi_0\bigl(\Diff\bigr)\,.
\end{equation}
Would we then address as physical \emph{observables} only those functions 
on phase space which are invariant under $\MCG$? The answer to this 
questions may well depend on the specific context at hand. But since 
$\MCG$ is generically a non-abelian and infinite group, the different 
answers will have significant effect on the size and structure of the 
space of physical states and observables. A 2+1 dimensional model 
where this has been studied in some detain is presented 
in~\cite{GiuliniLouko:1995}. 

Reducing the configuration space $\Riem$ by the action of $\Diff$ 
leads to what is called `superspace'\footnote{This notion is 
independent to similarly sounding ones in the theory of 
supersymmetric field theories}:
\begin{equation}
\label{eq:DefSuperspace}
\Super=\Riem/\Diff\,.
\end{equation}
It can be given the structure of a stratified 
manifold~\cite{Fischer:1970}, where the nested singular sets 
are labeled by the isotropy groups of $\Diff$ (i.e. the singular 
sets are the geometries with non-trivial isotropy group and nested 
according to the dimensionality of the latter.)

There is a natural way to resolve the singularities of $\Super$
\cite{Fischer:1986}, which can be described as follows: 
pick a point $\infty\in\Sigma$ (we shall explain below why the 
point is given that name) and consider the following subgroups
of $\Diff$ that fix $\infty$ and frames at $\infty$ respectively:
\begin{subequations}
\label{DiffSub}
\begin{alignat}{2}
\label{DiffSub1}
& \DiffInf &&\,:=\,
\bigl\{\phi\in\Diff : \phi(\infty)=\infty\bigr\}\,,\\
\label{DiffSub2}
& \DiffF &&\,:=\,
\bigl\{\phi\in\DiffInf : 
\phi_*\vert_\infty=\text{id}\vert_{T_\infty(\Sigma)}\bigr\}\,.
\end{alignat}
\end{subequations}
The resolved Superspace, $\SuperRes$, is then isomorphic 
to\footnote{The isomorphism is non canonical since we had to 
select a point $\infty\in\Sigma$.} 
\begin{equation}
\label{eq:DefSuperspaceRes}
\SuperRes:=\Riem/\DiffF\,.
\end{equation}
The point is that $\DiffF$ acts freely on $\Riem$ due to the
fact that diffeomorphisms that fix the frames at one point 
cannot contain non-trivial isometries.\footnote{To see this,
assume $\phi\in\DiffF$ is an isometry of $(q,\Sigma)$, where 
$\Sigma$ is connected. The set of fixed points is clearly 
closed. It is also open, as one readily sees by using the 
exponential map.} This, as well as the appropriate slicing 
theorems for the surjection $\Riem\rightarrow\SuperRes$(which 
already holds for the action of $\Diff$, see \cite{Fischer:1970}
and references therein) then establish a manifold structure 
of $\SuperRes$. In fact, we have a principle bundle
\begin{equation}
\label{eq:RiemAsPbundle}
{\DiffF}\to/ >->/<300>^i\Riem\to/->>/<300>^p\SuperRes\,.
\end{equation} 
The contractibility of $\Riem$ (which is an open 
convex cone in a topological vector space) implies that 
$\Riem$ is a universal classifying bundle and  $\SuperRes$
a universal classifying space for the group $\DiffF$. 
It also implies, via the long exact homotopy sequence for 
(\ref{eq:RiemAsPbundle}), that 
\begin{equation}
\label{eq:HomotopyGroups}
\pi_n\bigl(\DiffF\bigr)\cong
\pi_{n+1}\bigl(\SuperRes\bigr)\qquad
\text{for}\quad n\geq 0\,.
\end{equation}
Recall that $\pi_0$ of a topological group $G$ is a group (this is 
not true for arbitrary topological spaces) which is defined by 
$G/G^0$, where $G^0$ is the identity component. Hence we have 
\begin{equation}
\label{eq:DefMCGF}
\MCGF:=\DiffF/\DiffFId=:
\pi_0\bigl(\DiffF\bigr)\cong\pi_1\bigl(\SuperRes\bigr)\,.
\end{equation}
In this way we recognize the mapping-class group for frame-fixing 
diffeomorphisms of $\Sigma$ as the fundamental group of the 
singularity-resolved $\Diff$--reduced configuration space of canonical 
gravity. Next to (\ref{eq:DefMCG}) and (\ref{eq:DefMCGF}) we also 
introduce the analogous mapping class groups for point-fixing 
diffeomorphisms: 
\begin{equation}
\label{eq:DefMCGInf}
\MCGInf:=\DiffInf/\DiffInfId=:\pi_0\bigl(\DiffInf\bigr)\,.
\end{equation}

\subsection{Topologically open Cauchy surfaces}
\label{sec:TopOpen}
So far we assumed $\Sigma$ to be closed. This is the case of interest 
in cosmology. However, in order to model isolated systems, one is 
interested in 3-manifolds $\Sigma'$ with at least one asymptotically 
flat end, where here we restrict to the case of one end only. The 
topological implication behind `asymptotical flatness' is simply the 
requirement that the one-point compactification  
$\Sigma=\Sigma'\cup\infty$ (here $\infty$ is the point added) be a 
manifold. This is equivalent to the existence of a compact set 
$K\subset\Sigma'$ such that $\Sigma'-K$ is homeomorphic to 
$S^2\times\mathbb{R}$ (i.e. $\mathbb{R}^3-\text{ball}$). 

The analytic expressions given in Section\,\ref{sec:Introduction}
made no reference to whether $\Sigma$ is open or closed. 
In particular, the constraints are still given by (\ref{eq:ConstEq}).
However, in the open case it is not true anymore that the dynamical 
evolution is entirely driven by the constraints, as in 
(\ref{eq:EvolEqHam}) and (\ref{eq:TotalHamiltonianClosed}). 
Rather, we still have (\ref{eq:EvolEqHam}) but must change 
(\ref{eq:TotalHamiltonianClosed}) to\footnote{Here we neglect other 
surface integrals that arise in the presence of gauge symmetries 
other than diffeomorphism invariance whenever globally charged states
are considered.}
\begin{alignat}{2}
H[q,p;\alpha,\beta;\phi]
=\lim_{R\rightarrow\infty}
\Bigg\{
&\int_{B_R}\alpha C_s(q,p;\phi)
&&\,+\,\int_{B_R}\beta\cdot C_v(q,p;\phi)
\nonumber\\
\label{eq:TotalHamiltonianOpen}
+&\int_{S_R}\mathcal{E}(\alpha;q,p)
&&\,+\,\int_{S_R}\mathcal{M}(\beta;q,p)
\Bigg\}\,.
\end{alignat}
Here $B_R$ is a sequence of compact sets, labeled by their `radius' 
$R$, so that $R'>R$ implies $B_{R'}\supset B_R$ and 
$\lim_{R\rightarrow\infty}B_R=\Sigma'$. $S_R$ is equal to the 
boundary $\partial B_R$, which we assume to be an at least piecewise 
differentiable embedded 2-manifold in $\Sigma'$. $\mathcal{E}$ and 
$\mathcal{M}$ are the fluxes for energy and linear momentum if 
asymptotically for large $R$ the lapse function $\alpha$ assumes the 
constant value $1$ and $\beta$ approaches a translational 
Killing vector. Correspondingly, if $\beta$ approaches a rotational 
Killing vector, we obtain the flux for angular momentum 
(see \cite{BeigMurchadha:1987} for the analytic expressions in 
case of pure gravity). Since the constraints (\ref{eq:Constraints})
must still be satisfied as part of Einstein's equations, we see that 
`on shell' the Hamiltonian (\ref{eq:TotalHamiltonianOpen}) is a sum 
of surface integrals. Note also that even though the surface integrals 
do not explicitly depend on the matter variables $\phi$, as indicated 
in (\ref{eq:TotalHamiltonianOpen}), there is an implicit dependence 
through the requirement that $(q,p,\phi)$ satisfy the constraints 
(\ref{eq:Constraints}). This must be so since these surface integrals 
represent the total energy and momentum of the system, including the 
contributions from the matter. 

Let us consider the surface integral associated with the 
spatial vector field $\beta$. It is given by 
\begin{equation}
\label{eq:BetaSurfInt}
P(q,p;\beta)\,:=\,2\,\lim_{R\rightarrow\infty}
\Bigg\{\int_{S_R}p(n^\flat,\beta^\flat)\Bigg\}
\end{equation}
where $n$ is the outward pointing normal of $S_R$ in $\Sigma'$ 
and $n^\flat:=q(n,\cdot)$ etc. It is precisely minus the 
surface integral that emerges by an integration by parts from
the second integral on the right hand side of 
(\ref{eq:TotalHamiltonianClosed}) and which obstructs functional 
differentiability with respect to $p$. The addition of 
(\ref{eq:BetaSurfInt}) to (\ref{eq:TotalHamiltonianClosed}) just 
leads to a cancellation of both surface integrals thereby restoring 
functional differentiability for non-decaying $\beta$. This is 
precisely what was done in (\ref{eq:TotalHamiltonianOpen}).   
Conversely, this shows that the constraint $C_v(\beta)$ 
(cf. (\ref{eq:SmearedConstraints})) only defines a Hamiltonian 
vector field if $\beta$ tends to zero at infinity. Hence the 
constraints $C_v$ only generate asymptotically trivial 
diffeomorphisms. The rate of this fall-off is 
of crucial importance for detailed analytical considerations, 
but is totally unimportant for the topological ideas we are going
to present. For our discussion it is sufficient to work with 
$\Sigma=\Sigma'\cup\infty$. In particular, the group of spatial 
diffeomorphisms generated by the constraints may again be identified 
with $\DiffFId$. This is true since we are only interested in homotopy 
invariants of the diffeomorphism 
group\footnote{\label{foot:HomotopyInv} We already alert to the fact 
that homotopy invariants of the groups of homeomorphisms or 
diffeomorphisms of a 3-manifold $\Sigma$ are topological invariants 
of $\Sigma$ but \emph{not} necessarily also homotopy invariants of 
$\Sigma$. We will come back to this below.} and the group of 
diffeomorphisms generated by the constraints is homotopy equivalent 
to $\DiffFId$, whatever the precise fall-off conditions for the 
fields on $\Sigma'$ are. Moreover, the full group of diffeomorphisms, 
$\text{Diff}(\Sigma')$,is homotopy equivalent to 
$\DiffInf$.\footnote{To see this one needs, in particular, to know 
that $\text{Diff}(S^2)$ is homotopy equivalent to 
$SO(3)$~\cite{Smale:1959}.}

To sum up, the configuration space topology in Hamiltonian General 
Relativity is determined by the topology of $\DiffF$, where $\Sigma$ 
is a closed 3-manifold. This is true in case the Cauchy surface is 
$\Sigma$ and also if the Cauchy surface is open with one regular end, 
in which case $\Sigma$ is its one-point compactification.  
In particular, the fundamental group of configuration space is 
isomorphic to the mapping-class groups (\ref{eq:DefMCGF}). This is 
the object we shall now focus attention on. It has an obvious interest 
for the quantization program: for example, it is well known from 
elementary Quantum Mechanics that the inequivalent irreducible unitary 
representations of the 
fundamental group of the classical configuration space (the domain of the 
Schr\"odinger function) label inequivalent quantum sectors; see e.g. 
\cite{Giulini:1995b} and references therein. Even though in field theory it 
is not true that the classical configuration space is the proper 
functional-analytic domain of the Schr\"odinger state-functional, it remains 
true that its fundamental group---$\MCGF$ in our case---acts as group 
of (gauge) symmetries on the space of quantum states. Hence one is 
naturally interested in the structure and the representations of such 
groups. Some applications of these concepts in quantum cosmology 
and 2+1 quantum gravity may be found in \cite{GiuliniLouko:1992} 
\cite{GiuliniLouko:1995} respectively. The whole discussion on 
$\theta$-sectors\footnote{$\theta$ symbolically stands for the 
parameters that label the equivalence classes of irreducible unitary 
representations. This terminology is borrowed from QCD, where 
the analog of $\DiffF$ is the group of asymptotically trivial 
$SU(3)$ gauge-transformations, whose associated group of connected 
components---the analog of $\DiffF/\DiffFId$---is isomorphic to 
$\pi_3(SU(3))\cong\mathbb{Z}$. The circle-valued parameter $\theta$ 
then just labels the equivalence classes of irreducible unitary 
representations of that $\mathbb{Z}$.} in quantum gravity started 
in 1980 with the seminal paper \cite{FriedmanSorkin:1980} by John 
Friedman and Rafael Sorkin on the possibility of spin-1/2 states in 
gravity. We will discuss this below. Further discussion 
of mapping-class groups as symmetry groups in canonical 
quantum gravity and their physical relevance are given in 
\cite{Isham:1982,Sorkin:1986,Sorkin:1989,Aneziris-etal:1989b}. 

In analogy to standard gauge theories of Yang-Mills type, one may 
speculate that the higher homotopy groups of $\DiffF$ are also of 
physical significance, e.g. concerning the question of various 
types of 
anomalies~\cite{Alvarez-GaumeGinsparg:1985,NelsonAlvarez-Gaume:1985}. 
Such groups may be calculated for large classes of prime  
3-manifolds~\cite{Giulini:1995a} (the concept of a prime manifold is 
explained below), but not much seems to be known in the general 
reducible case.\footnote{For example, for connected sums of three or 
more prime manifolds, the fundamental group of the group of 
diffeomorphisms is not finitely 
generated~\cite{KalliongisMcCullough:1996}.} 

As already pointed out (cf. footnote\,\ref{foot:HomotopyInv}), the 
homotopy invariants of $\DiffF$ are topological but not necessarily 
also homotopy invariants of $\Sigma$ (cf. \cite{McCarty:1963}). 
For example, if $\Sigma$ is a spherical 
space form, that is $\Sigma\cong S^3/G$ with 
finite $G\subset SO(4)$, the mapping class group $\MCGF$ often fully
characterizes $\Sigma$ and can even sometimes distinguish two non-homeomorphic 
$\Sigma$'s which are homotopy equivalent. The latter happens in case of 
lens spaces, $L(p{,}q)$, where generally $p$ and $q$ denote any pair 
of coprime integers. Their mapping-class groups\footnote{A special 
feature of lens spaces is that $\MCG\cong\MCGInf\cong\MCGF$, where 
$\Sigma=L(p{,}q)$. These groups are also isomorphic to 
$\text{Isom}(\Sigma)/\text{Isom}^0(\Sigma)$, where $\text{Isom}$ 
denotes the group of isometries with respect to the metric of constant 
positive curvature. The property 
$\MCG\cong\text{Isom}(\Sigma)/\text{Isom}^0(\Sigma)$ is known to hold 
for many of the spherical space forms; an overview is given in 
\cite{Giulini:1994a} (see the table on p.\,922). It is a weakened form 
of the \emph{Hatcher Conjecture}~\cite{Hatcher:1978}, which states that the 
inclusion of $\text{Isom}(\Sigma)$ into $\Diff$ is a homotopy 
equivalence for all spherical space forms $\Sigma$. The Hatcher 
conjecture generalizes the Smale conjecture~\cite{Smale:1959}
(proven by Hatcher in \cite{Hatcher:1983}), to which it reduces for 
$\Sigma=S^3$.} are as follows: 
1.)~$\MCGF\cong\mathbb{Z}_2\times\mathbb{Z}_2$ if 
$q^2=1\,(\text{mod}\,p)$ and $q\ne\pm 1\,(\text{mod}\,p)$,
2.)~ $\MCGF\cong\mathbb{Z}_4$ if $q^2=-1\,(\text{mod}\,p)$,
and 3.)~$\MCGF\cong\mathbb{Z}_2$ in the remaining cases;
see Table~II. p.\,581 in \cite{Witt:1986b}. On the other hand, 
it is known that two lens spaces  $L(p{,}q)$ and $L(p{,}q')$
are homeomorphic iff\footnote{Throughout we use `iff' for `if and 
only if'.} $q'=\pm q\,(\text{mod}\,p)$ or 
$qq'=\pm 1\,(\text{mod}\,p)$~\cite{Reidemeister:1935} and homotopy 
equivalent iff $qq'$ or $-qq'$ is a quadratic residue $\text{mod}\,p$, 
i.e. iff $qq'=\pm n^2\,(\text{mod}\,p)$ for some integer 
$n$~\cite{Whitehead:1941}. For example, this implies that 
$L(15{,}1)$ and $L(15{,}4)$ are homotopy equivalent but not 
homeomorphic and that the mapping class group 
of $L(15{,}1)$ is $\mathbb{Z}_2$ whereas that of 
$L(15{,}4)$ is $\mathbb{Z}_2\times\mathbb{Z}_2$. Further distinctions 
can be made using the fundamental group of $\DiffF$. See the table 
on p.\,922 in \cite{Giulini:1994a} for more information.

\section{3-Manifolds}
\label{sec:3-Manifolds}
It is well known (e.g. \cite{Witt:1986a}) that Einstein's equations 
(i.e. the constraints) pose no topological obstruction to $\Sigma$.
Hence our $\Sigma$ can be any closed 3-manifold. For simplicity 
(and no other reason) we shall exclude non-orientable manifolds
and shall from now on simply say `3-manifold' if we mean 
\emph{closed oriented} 3-manifold.  

The main idea of understanding a general 3-manifold is to decompose 
it into simpler pieces by cutting it along embedded surfaces. 
Of most interest for us is the case where one cuts along 2-spheres, which 
results in the so-called \emph{prime decomposition}. The inverse process, 
where two 3-manifolds are glued together by removing
an embedded 3-disc in each of them and then identifying the remaining 
2-sphere boundaries in an orientation reversing (with respect to 
their induced orientations) way is called \emph{connected sum}. This is 
a well defined operation in the sense that the result is independent 
(up to homeomorphisms) of 1.)~how the embedded 3-discs where chosen and 
2.)~what (orientation reversing) homeomorphism between 2-spheres is 
used for boundary identification (this is nicely discussed in 
\S\,10 of \cite{BroeckerJaenich:DiffTop}). We write 
$\Sigma_1\uplus\Sigma_2$ to denote the connected sum of $\Sigma_1$ with 
$\Sigma_2$. The connected sum of a 3-manifold $\Sigma$ with a 3-sphere, 
$S^3$, is clearly homeomorphic to $\Sigma$. 

Let us now introduce some important facts and notation. The classic 
source is \cite{Hempel:ThreeManifolds}, but we also wish to mention 
the beautiful presentation in \cite{Hatcher:3-manifolds}. 
$\Sigma$ is called 
\emph{prime} if $\Sigma=\Sigma_1\uplus\Sigma_2$ implies that $\Sigma_1$ 
or $\Sigma_2$ is $S^3$. $\Sigma$ is called \emph{irreducible} if every 
embedded 2-sphere bounds a 3-disk. Irreducibility implies primeness 
and the converse is almost true, the only exception being the 
\emph{handle}, $S^1\times S^2$, which is prime but clearly not irreducible 
(no $p\times S^2$ bounds a 3-disc). Irreducible prime manifolds have  
vanishing second fundamental group.\footnote{Note that this is 
\emph{not} obvious from the definition of irreducibility, since 
non-zero elements of $\pi_2$ need not be representable as \emph{embedded} 
2-spheres. However, the \emph{sphere theorem} for 3-manifolds 
(see Thm.\,4.3 in \cite{Hempel:ThreeManifolds}) implies that at 
least \emph{some} non-zero element in $\pi_2$ can be so represented 
if $\pi_2$ is non-trivial.} The converse is true if every embedded 
2-sphere that bounds a homotopy 3-disk also bounds a proper 3-disk;
in other words, if fake 3-disks do not exist, which is equivalent 
to the Poincar\'e conjecture. So, if the Poincar\'e conjecture holds, 
a (closed orientable) 3-manifold $P$ is prime iff either $\pi_2(P)=0$ 
or $P=S^1\times S^2$. 

Many examples of irreducible 3-manifolds are provided by space forms, 
that is, manifolds which carry a metric of constant sectional 
curvature. These manifolds are covered by either $S^3$ or 
$\mathbb{R}^3$ and hence have trivial $\pi_2$. 
\begin{itemize}
\item
Space forms of positive curvature (also called `spherical space forms') 
are of the form $S^3/G$, where 
$G$ is a finite freely acting subgroup of $SO(4)$. Next to the cyclic 
groups $\mathbb{Z}_p$ these $G$ e.g. include the $SU(2)$ double 
covers of the symmetry groups of $n$-prisms, the tetrahedron, the 
octahedron, and the icosahedron, as well as direct products of those 
with cyclic groups of relatively prime (coprime) 
order.\footnote{As far as I am aware, it is still an open 
conjecture that spherical space forms comprise all primes with 
finite fundamental group, even given the validity of the Poincar\'e 
conjecture. In other words, it is only conjectured that 3-manifolds 
covered by $S^3$ are of the form $S^3/G$ where $G\subset SO(4)$ 
acting in the standard linear fashion. In \cite{Milnor:1957} Milnor 
classified all finite groups that satisfied some necessary condition 
for having a free action on $S^3$. The validity of the Smale 
conjecture \cite{Hatcher:1983} (which states that the embedding of 
$O(4)$ into $\text{Diff}(S^3)$ is a homotopy equivalence) eliminates 
those groups from the list which are not subgroups of $SO(4)$ 
\cite{Hatcher:1978}. What remains to be shown is that these groups 
do not admit inequivalent (equivalence being conjugation by some 
diffeomorphism) actions. The undecided cases are some cyclic 
groups of odd order; see \cite{Thomas:EllipticStructures}.} 
For example, the \emph{lens spaces} $L(p{,}q)$, where $q$ is coprime to 
$p$, are obtained by letting the generator of $\mathbb{Z}_p$ act on 
$S^3=\{(z_1,z_2)\in\mathbb{C}^2 : \vert z_1\vert^2+\vert z_2\vert^2=1\}$ 
by $(z_1,z_2)\mapsto (rz_1,r^qz_2)$ with $r=\exp(2\pi\mathrm{i}/p)$.
See e.g. \cite{Witt:1986b} for explicit presentations of the
groups $G\subset SO(4)$. 
\item
The flat space forms are of the form $\mathbb{R}^3/G$ where 
$\mathbb{R}^3$ carries the Euclidean metric and $G$ is a freely,
properly-discontinuously acting subgroups of the group 
$E_3=\mathbb{R}^3\rtimes O(3)$ of Euclidean motions. There are six 
such groups leading to orientable compact quotients 
(see~\cite{Wolf:SpacesConstCurv}, Thm.\,3.5.5 and Cor.\,3.5.10): 
the lattice $\mathbb{Z}^3\subset\mathbb{R}^3$ of translations 
and five finite downward extensions\footnote{\label{foot:Extensions}
Let $G$ be a group with normal subgroup $N$ and quotient $G/N=Q$. 
We call $G$ either an \emph{upward extension} of $N$ by $Q$ or a 
\emph{downward extension} of $Q$ by $N$; see \cite{Atlas:FiniteGroups}, 
p.\,XX.} of it by $\mathbb{Z}_2$, $\mathbb{Z}_3$, $\mathbb{Z}_4$, 
$\mathbb{Z}_6$, and $\mathbb{Z}_2\times\mathbb{Z}_2$. These gives 
rise to the 3-torus $T^3$ and five spaces regularly covered by it. 
\item
Space forms of negative curvature (also called `hyperbolic space forms') 
are given by $H^3/G$, where $H^3=\{(t,\vec x)\in\mathbb{R}^{(1,3)} :
t=\sqrt{\vec x^2+1}\}$ is the hyperbola in Minkowski space and $G$ is a  
freely, properly-discontinuously acting subgroup of the Lorentz group
$O(1,3)$ leading to orientable compact quotients. They are much harder to 
characterize explicitly.
\end{itemize}

Flat and hyperbolic space forms are covered by $\mathbb{R}^3$ so 
that all their homotopy groups higher than the first are trivial. 
The class of topological spaces for which $\pi_k=\{0\}$ for all 
$k>1$ is generally called $K(\pi,1)$ (Eilenberg MacLane spaces of 
type $(\pi,1)$). In a sense, most primes are $K(\pi,1)$ and much  
remains to be understood about them in general. Considerably more 
is known about a special subclass, the so called \emph{sufficiently large} 
$K(\pi,1)$ 3-manifolds or \emph{Haken manifolds}. They are characterized 
by the property that they contain an embedded \emph{incompressible} 
Riemann surface $R_g$, i.e. that if $e:R_g\rightarrow\Pi$ is the 
embedding then $e_*:\pi_1(R_g)\rightarrow\pi_1(\Pi)$ is 
injective.\footnote{In other words, every loop on $e(R_g)\subset\Pi$ 
that bounds a 2-disc in $\Pi$ (and is hence contractible in $\Pi$) 
also bounds a 2-disc on $i(R_g)$ (an is hence contractible in $e(R_g)$).} 
Simple examples are provided by the products $S^1\times R_g$. 
An important conjecture in 3-manifold theory states 
that every irreducible 3-manifold with infinite fundamental group 
is virtually Haken, that is, finitely covered by a Haken manifold. 
If this is the case, any prime with infinite fundamental group 
allows for an immersion $e:R_g\rightarrow\Pi$ such that 
$e_*:\pi_1(R_g)\rightarrow\pi_1(\Pi)$ is injective.  

\begin{figure}[ht]
\begin{minipage}[c]{0.35\linewidth}
\centering\epsfig{figure=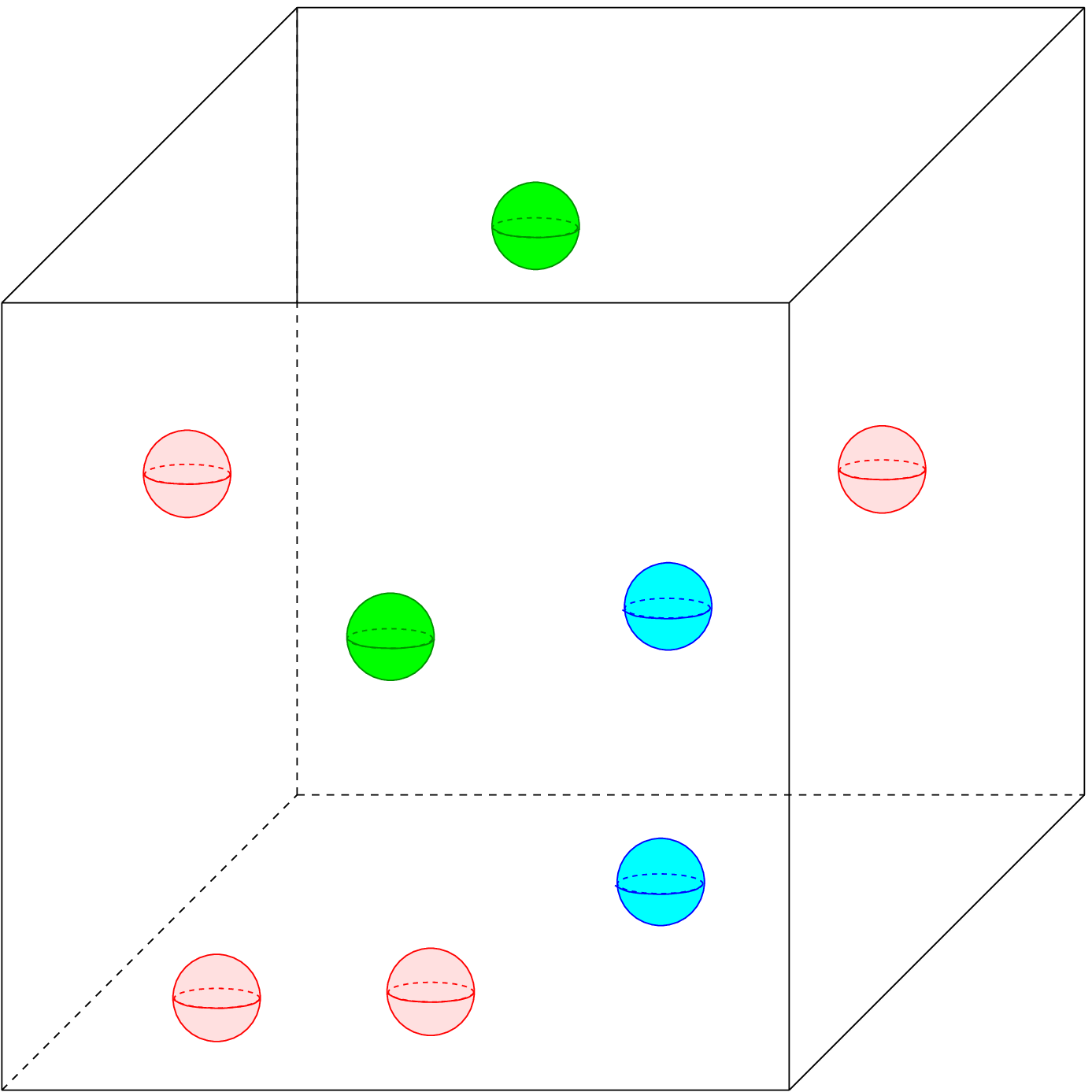,width=1.0\linewidth}
\end{minipage}
\hfill
\begin{minipage}[c]{0.60\linewidth}
\centering\epsfig{figure=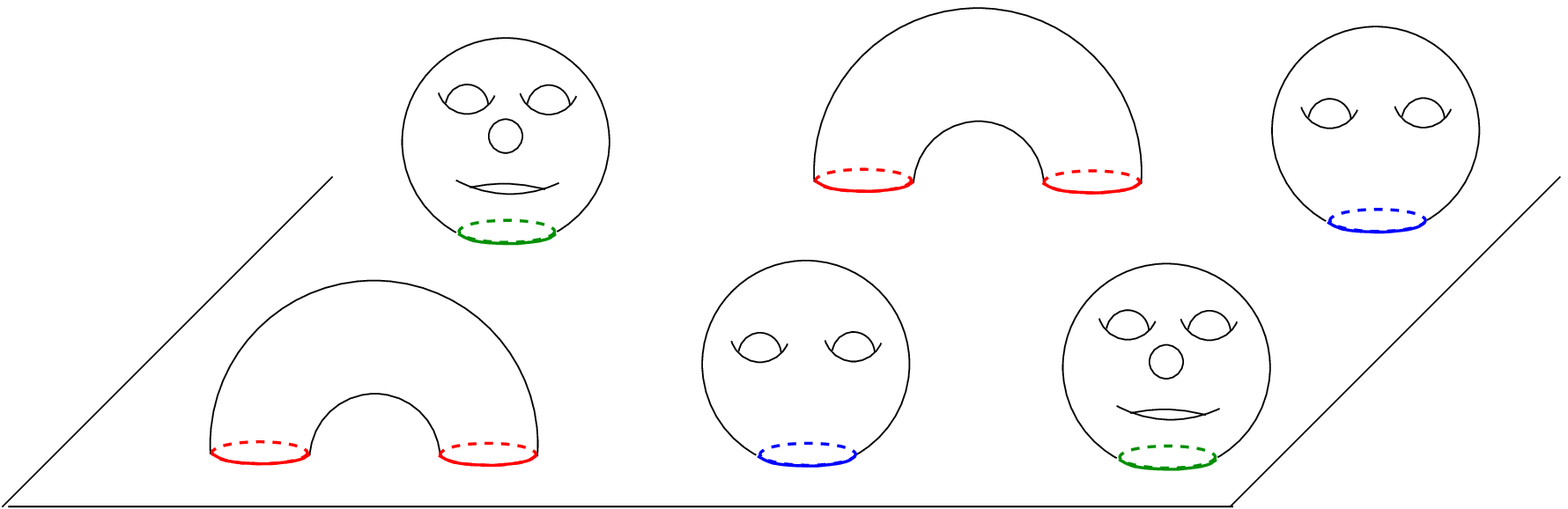,width=1.0\linewidth}
\end{minipage}
\put(-84,27){\tiny $H_1$}
\put(-167,-11){\tiny $H_2$}
\put(-39,-10){\tiny $\Pi_1$}
\put(-148,36){\tiny $\Pi_2$}
\put(-29,38){\tiny $\Pi_3$}
\put(-88,-10){\tiny $\Pi_4$}
\put(-259,20){\tiny $\sigma'_1$}
\put(-339,19){\tiny $\sigma_1$}
\put(-310,-40){\tiny $\sigma'_2$}
\put(-333,-41){\tiny $\sigma_2$}
\put(-284,4){\tiny $S_3$}
\put(-283,-46){\tiny $S_4$}
\put(-315,0){\tiny $S_1$}
\put(-299,47){\tiny $S_2$}
\caption{\label{fig:ConnectedSum}%
The connected sum of two handles $H_{1{,}2}$ and four 
irreducible primes $\Pi_1,\cdots\Pi_4$, with $\Pi_1$ diffeomorphic 
to $\Pi_2$ and $\Pi_3$ diffeomorphic to $\Pi_4$. For later 
application concerning the mapping class groups it is advisable 
to represent a handle as a cylinder $[0,1]\times S^3$ with both 
ends, $0\times S^2$ and $1\times S^2$, separately 
attached by a connecting 2-sphere. The connecting spheres are 
denoted by $\sigma_i,\sigma'_i$ for $H_i$ ($i=1,2$) and $S_i$ for 
$\Pi_i$ ($i=1,\cdots,4$).
The left picture gives an \emph{internal view}, in which only the connecting 
spheres are seen (and not what is behind), the right picture gives an 
\emph{external view} from three dimensions onto a two-dimensional 
analogous situation that also reveals the topological structures 
behind the connecting spheres.}
\end{figure}

Now, given a connected sum ($N=n+m$)
\begin{equation}
\label{eq:ConnSum}
\Sigma
=\biguplus_{i=1}^{N}P_i
=\left\{\biguplus_{i=1}^{n}\Pi_i\right\}\uplus
 \left\{\biguplus^{m}S^1\times S^2\right\}\,,
\end{equation}
where notationally we distinguish between unspecified primes, denoted 
by $P_i$, and irreducible primes (i.e. those different from $S^1\times S^2$),
denoted by $\Pi_i$, so that $P_i=\Pi_i$ for $1\leq i\leq n$ and 
$P_i=S^1\times S^2$ for $n<i\leq N$. A simple application of 
van Kampen's rule gives that the fundamental group of the connected sum 
in 3 (and higher) dimensions is isomorphism to the free product 
(denoted by~$*$) of the fundamental groups the factors (since the 
connecting spheres are simply connected):      
\begin{equation}
\label{eq:FundGroupConnSum}
\pi_1(\Sigma)=\pi_1(P_1)*\cdots *\pi_1(P_N)\,.
\end{equation} 
The converse is also true: a full decomposition of the group 
$\pi_1(\Sigma)$ into free products corresponds to a decomposition
of $\Sigma$ into the connected sum of primes (known as Kneser's 
conjecture).   

The existence of prime decompositions was first shown by 
Kneser~\cite{Kneser:1929}, the uniqueness (up to permutations 
of factors) by Milnor~\cite{Milnor:1962}. Regarding the latter we 
need to recall that we consider all manifolds to be oriented. 
Many orientable primes do not allow for orientation reversing self 
diffeomorphism; they are called \emph{chiral}. The table on p.\,922
of \cite{Giulini:1994a} lists which spherical and flat space 
forms are chiral; most of them are. Chiral primes with opposite 
orientation must therefore be considered as different prime 
manifolds. 

Irreducible primes can be further decomposed by cutting them along 
2-tori, which is the second major decomposition device in Thurston's 
\emph{Geometrization Program} of 3-manifolds. Here we will not 
enter into this.

\section{Mapping class groups}
\label{sec:MCG}
Mapping class groups can be studied through their action on the 
fundamental group. Consider the fundamental group of $\Sigma$ based
at $\infty\in\Sigma$. We write $\pi_1(\Sigma,\infty)$ or sometimes 
just $\pi_1$ for short. There are homomorphisms  
\begin{alignat}{3}
\label{eq:Map-hF}
& h_{\mathrm{F}}&&\,:\,\MCGF&&\rightarrow\text{Aut}(\pi_1)\,,\\
\label{eq:Map-hInf}
& h_{\infty}&&\,:\,\MCGInf&&\rightarrow\text{Aut}(\pi_1)\,,\\
\label{eq:Map-h}
& h&&\,:\,\MCG&&\rightarrow\text{Out}(\pi_1)\,,
\end{alignat}
where the first two are given by  
\begin{equation}
\label{eq:DefMap-both}
[\phi]\mapsto \bigl([\gamma]\mapsto[\phi\circ\gamma]\bigr)\,.
\end{equation}
Here $[\phi]$ denotes the class of $\phi\in\DiffF$ in 
$\DiffF/\DiffFId$ (or of $\phi\in\DiffInf$ in 
$\DiffInf/\DiffInfId$) and the other two square brackets the homotopy 
classes of the curves $\gamma$ and $\phi\circ\gamma$. As regards 
(\ref{eq:Map-h}), it is not difficult to see that in $\Diff$
any inner automorphism of $\pi_1(\Sigma,\infty)$ can be generated 
by a diffeomorphism that is connected to the identity (in $\Diff$,
not in $\DiffInf$ or $\DiffF$). Hence we have to factor out the 
inner automorphisms for the map $h$ to account for the possibility 
to move the basepoint $\infty$. More precise arguments are given 
in Section\,3 of \cite{Giulini:1997a}.

\begin{figure}[htb]
\begin{minipage}[c]{0.40\linewidth}
\centering\epsfig{figure=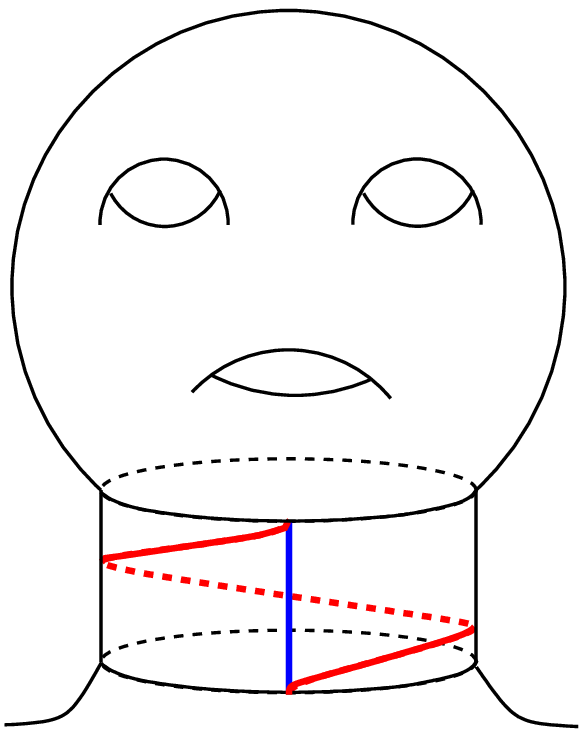,width=0.9\linewidth}
\end{minipage}
\hfill
\begin{minipage}[c]{0.58\linewidth}
\centering\epsfig{figure=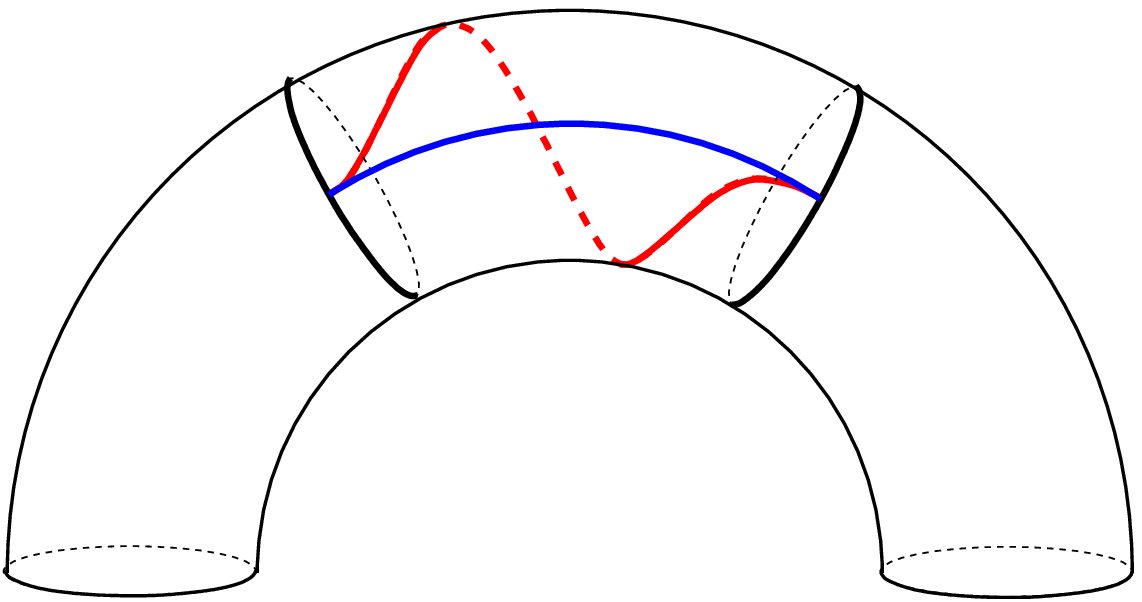,width=0.9\linewidth}
\end{minipage}
\put(-64,10){\footnotesize $S_1$}
\put(-150,10){\footnotesize $S_2$}
\put(-290,60){$\Sigma$}
\put(-310,-26){\footnotesize $S_1$}
\put(-310,-64){\footnotesize $S_2$}
\caption{\label{fig:DehnTwists}%
Both pictures show rotations parallel to spheres $S_1$ and $S_2$. 
On the left, a rotation of the manifold $\Sigma$ parallel to 
spheres both centered at $\infty$, or, if $\Sigma=\Pi$ is a 
prime manifold in a connected sum, parallel to the connecting 
2-sphere. On the right is a rotation parallel to two meridian 
spheres in a handle. The support of the diffeomorphism is on 
the cylinder bound by $S_1$ and $S_2$. In either case its 
effect is depicted by the two curves connecting the two 
spheres: the diffeomorphism maps the straight to the curved 
line. This 2-dimensional representation is  deceptive insofar 
as here the two straight and the curved lines are not 
homotopic (due to $\pi_1(S^1)=\mathbb{Z}$), whereas 
in 3 dimensions they are (due to the triviality of $\pi_1(S^2)$).
}
\end{figure}

The images of $h_\infty$ and $h_{\mathrm{F}}$ coincide but their 
domains may differ, i.e. the groups $\MCGF$ and $\MCGInf$ are not 
necessarily isomorphic. Let us explain this a little further. 
Consider the fibration 
\begin{equation}
\label{eq:Fibration-DiffInf:DiffF}
{\DiffF}\to/ >->/<300>^i\DiffInf\to/->>/<300>^p\text{GL}^+(3{,}\mathbb{R})\,,
\end{equation} 
where $p(\phi):=\phi_*\vert_{\infty}$ (here we identify 
$GL^+(3{,}\mathbb{R})$ with the orientation preserving linear 
isomorphisms of $T_\infty(\Sigma)$). Associated with this fibration 
is a long exact sequence of homotopy groups,  which ends with
\begin{equation}
\label{eq:ExSeq-DiffInf:DiffF}
1\to\pi_1(\DiffF)\to\pi_1(\DiffInf)\to^{p_*}\mathbb{Z}_2
\to^{\partial_*}\MCGF\to^{i_*}\MCGInf\to 1
\end{equation} 
where the leftmost zero comes from $0=\pi_2(\text{GL}^+(3{,}\mathbb{R}))$
and the $\mathbb{Z}_2$ in the middle is 
$\pi_1(\text{GL}^+(3{,}\mathbb{R}))$. Now, there are only two 
possibilities as regards the image of $p_*$:
\begin{itemize}
\item[1.]
$\text{Image}(p_*)=\mathbb{Z}_2=\text{kernel}(\partial_*)
\Rightarrow\MCGF\cong\MCGInf$.
\item[2.]
$\text{Image}(p_*)=\{0\}=\text{kernel}(\partial_*)
\Rightarrow$ $\MCGF$ is a downward extension 
(recall footnote\,\ref{foot:Extensions}) of $\MCGInf$ by 
$\mathbb{Z}_2$. 
\end{itemize}
Let us focus on the second possibility. We first note that 
$\text{image}(\partial_*)$ lies in the kernel of $h_{\mathrm{F}}$.
This is true because the images of $h_{\mathrm{F}}$ and $h_\infty$
coincide, so that   
$h_{\mathrm{F}}\circ\partial_*=h_\infty\circ i_*\circ\partial_*$,
which is the trivial map onto the identity in $\text{Aut}(\pi_1)$,
since by exactness $i_*\circ\partial_*$ is the trivial map.    
The diffeomorphism that represents the non-trivial image of 
$\partial_*$ can be represented by a rotation parallel to two 
concentric small spheres centered at $\infty$; see the left picture 
in Fig.\,\ref{fig:DehnTwists}. From this picture it also becomes clear 
that the diffeomorphism representing the $2\pi$-rotation can be 
chosen of disjoint support from those diffeomorphisms representing 
other elements of $\MCGF$. Hence, if  $\MCGF$ is a $\mathbb{Z}_2$ 
extension of $\MCGInf$, it is \emph{central}. 
  
Manifolds for which $\MCGF$ is a (downward) central $\mathbb{Z}_2$ 
extension of $\MCGInf$ are called \emph{spinorial}. If they are used to 
model isolated systems (by removing $\infty$ and making the end 
asymptotically flat) their asymptotic symmetry group is not the 
Poincar\'e group (as discussed in \cite{BeigMurchadha:1987}) but 
its double cover (i.e. the identity component of the homogeneous 
symmetry group is $SL(2{,}\mathbb{C})$ rather than the proper 
orthochronous Lorentz group). The origin of this is purely 
topological and has nothing to do with quantum theory, though the 
possible implications for quantum gravity are particularly 
striking, as was first pointed out in a beautiful paper by 
Friedman and Sorkin \cite{FriedmanSorkin:1980}: it could open 
up the possibility to have half-integer spin states in pure 
gravity.\footnote{Topologically speaking, this is somewhat 
analogous to the similar mechanism in the Skyrme 
model~\cite{Skyrme:1971}, where loops in configuration space 
generated by $2\pi$-rotations are non-contractible iff the 
skyrmion's winding number (its baryon number) is 
odd~\cite{Giulini:1993}). The analogy to 
the mechanism by which half-integer spin states can arise 
in gauge theories of integer spin 
fields~\cite{HasenfratzHooft:1976,Goldhaber:1976} is less 
close, as they need composite objects, e.g. from magnetic 
monopoles and electric charges.}

A connected sum  is spinorial iff it contains at least one 
spinorial prime~\cite{Hendriks:1977}. Except for the lens spaces 
and handles all primes are spinorial, hence a 3-manifold is 
non-spinorial iff it is the connected sum of lens spaces and 
handles. Let us digress a little to explain this in somewhat 
more detail. 

\subsection*{A small digression on spinoriality}
That lens spaces and handles are non-spinorial is easily 
visualized. Just represent them in the usual fashion by embedding 
a lens or the cylinder $S^2\times[0,1]$ in $\mathbb{R}^3$, 
with the boundary identifications understood. Place the base 
point, $\infty$, on the vertical symmetry axis and observe that 
the rotation around this axis is compatible with the boundary 
identifications and therefore defines a diffeomorphism of the 
quotient space. A rotation parallel to two small spheres centered 
at $\infty$ can be continuously undone by rotating the body in 
$\mathbb{R}^3$ and keeping a neighborhood of $\infty$ fixed. 
This visualization also works for arbitrary connected sums of 
lens spaces and handles. 

Spinoriality is much harder to prove. The following theorem has 
been shown by Hendriks (\cite{Hendriks:1977}, Thm.\,1 in \S\,4.3), 
and later in a more constructive fashion by Plotnick 
(\cite{Plotnick:1986}, Thm.\,7.4):
\begin{thm}
\label{thm:Hendriks}
Let $\Sigma$ be a closed (possibly non-orientable) 3-manifold and $
\Sigma':=\Sigma-B_3$, where $B_3$ is an open 3-disc. A $2\pi$-rotation 
in $\Sigma'$ parallel to the boundary 2-sphere 
$\partial\Sigma'$ is homotopic to $\mathrm{id}_{\Sigma'}$ 
rel.\,$\partial\Sigma'$ (i.e. fixing the boundary throughout) 
iff every prime summands of $\Sigma$ is taken from the following 
list:  
\begin{itemize}
\item[1.)]
$S^3_h/G$, where $S^3_h$ is a homotopy sphere and $G$ a finite freely 
acting group all Sylow subgroups of which are cyclic,
\item[2.)]
the handle $S^1\times S^2$,
\item[3.)]
the (unique) non-orientable handle $S^1\tilde\times S^2$,
\item[4.)]
$S^1\times\RP^3$, where $\RP^3$ denotes 3-dimensional real 
projective space.
\end{itemize}
\end{thm}
Since here we excluded non-orientable manifolds from our discussion, 
we are not interested in 3.) and 4.). Clearly, $S^3$ is the only 
homotopy 3-sphere if the Poincar\'e conjecture holds. Of the remaining 
spherical space forms $S^3/G$ the following have cyclic Sylow 
subgroups\footnote{Each of the other groups contains as subgroup 
the `quaternion group' $D_8^*:=\{\pm 1,\pm i,\pm j,\pm k\}$, which 
is non abelian and of order $8=2^3$. Hence their 2-Sylow subgroups are 
not cyclic.}: 
\begin{itemize}
\item[a.)]
$G=\mathbb{Z}_p$ (giving rise to the lens spaces),
\item[b.)] 
$G=D^*_{4m}\times\mathbb{Z}_p$ for $m=$ odd and $4m$ coprime to 
$p\geq 0$. 
Here $D^*_{4m}$ is the $SU(2)$ double cover of $D_{2m}\subset SO(3)$,
the order $2m$ symmetry group of the $m$--prism.  
\item[c.)]
$G=D'_{2^km}\times\mathbb{Z}_p$ for $m=$ odd, $k>3$, and $2^km$ 
coprime to $p\geq 0$. Here $D'_{2^km}$ is a (downward) central 
extension of $D^*_{4m}$ by $\mathbb{Z}_{2^{k-2}}$.\footnote{%
We have $D_{2m}=\langle \alpha,\beta:\alpha\beta=\beta\alpha^{-1},\,
\alpha^m=\beta^2=1\rangle$, where $\alpha$ is a $2\pi/m$--rotation 
of the $m$-prism (vertical axis) and $\beta$ is a $\pi$ rotation 
about a horizontal axis. Then 
$D^*_{4m}=\langle \alpha,\beta:\alpha\beta=\beta\alpha^{-1},\,
\alpha^m=\beta^2\rangle=\langle a,b:ab=ba^{-1},\,a^m=b^4=1\rangle$,
where $a:=\alpha\beta^2$ and $b:=\beta$ (to show equivalence of 
these two presentations one needs that $m$ is odd), 
and $D'_{2^km}=\langle A,B:AB=BA^{-1},\,A^m=B^{2^k}=1\rangle$. 
The center of $D'_{2^km}$ is generated by $B^2$ and isomorphic to 
$\mathbb{Z}_{2^{k-1}}$. $B^4$ generates a central subgroup 
$\langle B^4\rangle$ isomorphic to $\mathbb{Z}_{2^{k-2}}$ and 
$D'_{2^km}/\langle B^4\rangle\cong D^*_{4m}$.}
\end{itemize}

Now there is a subtle point to be taken care of: that a 
diffeomorphism is \emph{homotopic} to the identity means that there 
is a one-parameter family of \emph{continuous} maps connecting it to the 
identity. This does not imply that it is \emph{isotopic} to the 
identity, which means that there is a one-parameter family of 
\emph{diffeomorphisms} connecting it to the identity. In case of the 
lens spaces it is easy to `see' the isotopy, as briefly explained 
above. However, in the cases b.) and c.) it was proven by Friedman 
\& Witt in \cite{FriedmanWitt:1986} that the homotopy ensured by 
Thm.\,\ref{thm:Hendriks} does not generalize to an isotopy,
so that these spaces again are spinorial. Taken together with 
Thm\,\ref{thm:Hendriks} this completes the proof of the statement 
that the only non-spinorial 3-manifolds are lens spaces, handles, and 
connected sums between them. 
    
\begin{figure}[htb]
\centering\epsfig{figure=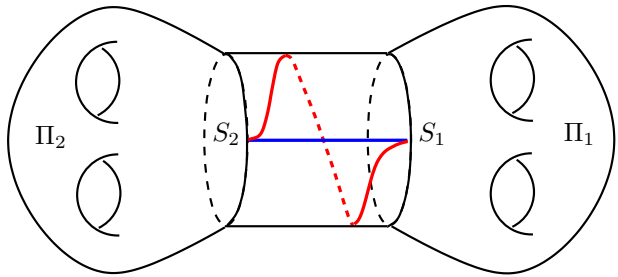,width=0.65\linewidth}
\put(-20,51){$\Pi_1$}
\put(-75,51){$S_1$}
\put(-153,51){$S_2$}
\put(-220,50){$\Pi_2$}
\caption{\label{fig:2PiTwist}%
The connected sum of two irreducible primes $\Pi_1$ and $\Pi_2$.
The relative $2\pi$-rotation is a diffeomorphism with support 
inside the cylinder bounded by the 2-spheres $S_1$ and $S_2$. 
It transforms the straight line connecting $S_1$ and $S_2$ 
into the curved line. If $\Pi_i=S^3/G_i$ ($i=1,2$), where 
$G_{1,2}\in SO(4)$ are taken from the families  
$D^*_{4m}\times\mathbb{Z}_p$ or $D'_{2^km}\times\mathbb{Z}_p$ 
mentioned under b.) and c.) in the text, this diffeomorphism is 
homotopic but not isotopic to the identity.}
\end{figure}

Note that this result also implies the existence of diffeomorphisms
in $\Diff$ which are homotopic but not isotopic to the identity. 
For example, take the connected sum $\Sigma=\Pi_1\uplus\Pi_2$ of two 
primes listed under b.) or c.). The $2\pi$-rotation parallel 
to the connecting 2-sphere will now be an element of $\Diff$ that 
is homotopic but not isotopic to the 
identity~\cite{FriedmanWitt:1986}; see Fig.\,\ref{fig:2PiTwist}.
This provides the first known example of such a behavior in 3-dimensions
(in two dimensions it is known not to occur), though no example is known 
where this happens for a prime 3-manifold. In fact, that homotopy implies 
isotopy has been proven for a very large class of primes, including all 
spherical space forms, the handle $S^1\times S^2$, Haken manifolds and 
many non-Haken $K(\pi,1)$ (those which are Seifert fibered). 
See e.g. Thm.\,A1 of \cite{Giulini:1995a} for a list of references.

\subsection*{General Diffeomorphisms}
It can be shown that if all primes in a connected sum satisfy the 
homotopy-implies-isotopy property, the kernel of $h_\mathrm{F}$ is 
isomorphic to $\mathbb{Z}_2^{m+n_s}$, where $m$ is the number of 
handles and $n_s$ the number of spinorial primes.\footnote{%
This follows from Thm.\,1.5 in \cite{McCullough:1990} together 
with the fact that for a manifold $\Pi$ with vanishing $\pi_2$ 
two self-diffeomorphisms $\phi_{1,2}$ are homotopic if their 
associated maps 
$h_\infty:[\phi_{1,2}]\mapsto\mathrm{Aut}(\pi_1(\Pi,\infty))$
coincide.} This group is 
generated by the diffeomorphisms depicted in Fig.\,\ref{fig:DehnTwists},
one neck-twist (left picture) for each spinorial primes and one 
handle-twist (right picture) for each handle. It can also be shown 
that the mapping class groups of each prime injects into the mapping 
class groups of the connected sum in which it 
occurs~\cite{HendriksMcCullough:1987}. This means that a 
diffeomorphism that has support in a prime factor $P$ (we call such 
diffeomorphisms \emph{internal}) and is not isotopic to the identity 
within in the space of all diffeomorphisms fixing the connecting 
2-sphere is still not isotopic to the identity in $\DiffF$. 
This statement would be false if $\DiffF$ were replaced by $\Diff$.
We will briefly come back to this point at the end of this section. 

In analogy to (\ref{eq:Map-hF}), for each prime $P$, there is a map 
$h_{\mathrm{F}}:\mathcal{G}_{\mathrm{F}}(P)\rightarrow\mathrm{Aut}(\pi_1(P))$
which is (almost) surjective in many cases. For example, if 
$\Pi$ is Haken, $h_{\mathrm{F}}$ maps onto 
$\mathrm{Aut}^+(\pi_1(\Pi))$~\cite{Hatcher:1976}, the subgroup of 
orientation preserving automorphisms. If $\Pi$ is not chiral, i.e. 
allows for orientation reversing self-diffeomorphisms, 
$\mathrm{Aut}^+(\pi_1(\Pi))\subset\mathrm{Aut}(\pi_1(\Pi))$ is a subgroup 
of index two (hence normal). However, if $\Pi$ is chiral, we have  
$\mathrm{Aut}^+(\pi_1(\Pi))=\mathrm{Aut}(\pi_1(\Pi))$ and hence surjectivity. 
Since Haken manifolds are all spinorial, we can now say that their 
mapping class group is a central $\mathbb{Z}_2$ extension of 
$\mathrm{Aut}^+(\pi_1(\Pi))$. 

For spherical space forms the mapping class groups have all been 
determined in \cite{Witt:1986b}. For $S^1\times S^2$ it is 
$\mathbb{Z}_2\times\mathbb{Z}_2$, where, say, the first 
$\mathbb{Z}_2$ is generated by the twist as depicted on the right 
in Fig.\,\ref{fig:DehnTwists}. The second $\mathbb{Z}_2$ corresponds 
to $\mathrm{Aut}(\pi_1(\Pi))=\mathrm{Aut}(\mathbb{Z})$ and is generated 
by a reflection in the circle $S^1$. If one thinks of a handle in a 
prime decomposition as a cylinder being attached with both ends, as 
depicted in Fig.\,\ref{fig:ConnectedSum}, the latter diffeomorphism 
corresponds to exchanging the two cylinder ends (in an orientation 
preserving fashion), which is sometimes called a \emph{spin} of a handle. 

Suppose now that we are given a general connected sum 
(\ref{eq:ConnSum}) and that we know the mapping class group 
of each prime in terms of a finite presentation (finitely many 
generators and relations). We can then determine a finite 
presentation of $\MCGF$ by means of the so-called Fouxe-Rabinovitch 
presentation for the automorphism group of a free product of 
groups developed in \cite{Fouxe-Rabinovitch:1940,Fouxe-Rabinovitch:1941}; 
see also \cite{McCulloughMiller:1986} and \cite{Gilbert:1987}. 
Let generally 
\begin{equation}
\label{eq:FreeProduct}
G=\underbrace{G_{(1)}*\cdots *G_{(n)}}_{%
\mathrm{each}\,\not\cong\,\mathbb{Z}}\,*\,
\underbrace{G_{(n+1)}*\cdots * G_{(n+m)}}_{\mathrm{each}\,\cong\,\mathbb{Z}}
\end{equation}
be a free product of groups corresponding to the decomposition 
(\ref{eq:ConnSum}). Let a set $\{g_{(i)1},\cdots, g_{(i)n_i}\}$ of 
generators for each $G_{(i)}$ be chosen. Clearly, for $n<i\leq n+m$
we have $n_i=1$ so that we also write $g_{(i)1}=:g_{(i)}$. 
The generators of $\mathrm{Aut}(G)$ can now be characterized by 
their action on these generators as follows (only non-trivial 
actions are listed): 
\begin{itemize}
\item[1.)]
The generators of each $\mathrm{Aut}(G_{(i)})$ for $1\leq i\leq n$.
As mapping-class generator these are called \emph{internal}. 
\item[2.)]
The $m$ generators $\sigma_i$ $(1\leq i\leq m)$ whose 
effect is $\sigma_i(g_{(n+i)})=g^{-1}_{(n+i)}$, i.e. generating 
$\mathrm{Aut}(\mathbb{Z})\cong\mathbb{Z}_2$. As mapping-class 
generator $\sigma_i$ is called a \emph{spin of the $i$-th handle}.
\item[3.)]
One generator $\omega_{(i)(k)}$ for each pair of distinct but 
isomorphic groups $G_{(i)},G_{(k)}$ ($1\leq i,k\leq n+m$), whose 
effect it is to slotwise exchange the sets  
$\{g_{(i)1},\cdots, g_{(i)n_i}\}$ and 
$\{g_{(k)1},\cdots, g_{(k)n_k}\}$. (Here, for $1\leq i,k\leq n$, 
we assume the generators of isomorphic groups to be chosen such that 
they correspond under a fiducial isomorphism, in particular 
$n_i=n_k$). As mapping-class generator $\omega_{(i)(k)}$ is 
called the \emph{exchange of prime $i$ with prime $k$.}  
\item[4.)]
One generator $\mu_{(i)j,(k)}$ for each $1\leq i\leq n+m$, 
$1\leq j\leq n_i$, and $1\leq k\leq n$, whose effect is 
to map each generator $g_{(k)l}$ ($1\leq l\leq n_k$) to 
$g^{-1}_{(i)j}\cdot g_{(k)l}\cdot g_{(i)j}$. 
As mapping-class generator $\mu_{(i)j,(k)}$ is called a 
\emph{slide of the (irreducible) prime $k$ through prime $i$ along $g_{(i)j}$.}
\item[5.)]
One pair of generators, $\lambda_{(i)j,(k)}$ and $\rho_{(i)j,(k)}$ 
for each $1\leq i\leq n+m$, $1\leq j\leq n_i$, 
and $n\leq k\leq n+m$. The effect of $\lambda_{(i)j,(k)}$ is to 
map each generator $g_{(k)}$ to $g^{-1}_{(i)j}\cdot g_{(k)}$
(i.e. left multiplication) and that of $\rho_{(i)j,(k)}$ to 
map each $g_{(k)}$ to $g_{(k)}\cdot g_{(i)j}$ (i.e. 
right multiplication). As mapping-class generator 
$\lambda_{(i)j,(k)}$ is called the \emph{slide of the left end of handle 
$k$ through prime $i$ along $g_{(i)j}$} and $\rho_{(i)j,(k)}$ is 
called the \emph{slide of the right end of handle $k$ through prime $i$ along 
$g_{(i)j}$.}
\end{itemize}
Of these generators the mapping class group realizes all those 
listed in 2.)-5.), but might leave out some in 1.) in case 
$h_{\mathrm{F}}:
\mathcal{G}_{\mathrm{F}}(\Pi_i)\rightarrow \mathrm{Aut}(\pi_1(\Pi_i))$
is not surjective for some $i\leq n$. In that case just replace 1. by the 
generators of the image of $h_{\mathrm{F}}$ (which might be a larger 
set than the generators of $\mathrm{Aut}(\pi_1(\Pi_i))$). Finally, we 
have to add the generators of the kernel of $h_{\mathrm{F}}$. As already 
stated, this kernel is given by the direct product of $n_s+m$ copies of 
$\mathbb{Z}_2$, where $n_s$ is the number of spinorial primes, if we 
assume the `homotopy-implies-isotopy'-property for all primes. In that 
case we have found all generators after adjoining these additional $n_s+m$ 
generators. A complete list of relations can then be found from the 
Fouxe-Rabinovitch relations for $\mathrm{Aut(G)=G_1*\cdots G_{n+m}}$ 
(see Chapter\,5.1 of 
\cite{McCulloughMiller:1986})\footnote{The original papers by 
Fouxe-Rabinovitch \cite{Fouxe-Rabinovitch:1941,Fouxe-Rabinovitch:1940}  
contained some errors in the relations which were corrected in
\cite{McCulloughMiller:1986}; see also \cite{Gilbert:1987}.}  
and some added relations which the $n_s+m$ added generators 
have to satisfy. The latter are not difficult to find due to 
the simple geometric interpretation of the diffeomorphisms
of Fig.\,\ref{fig:DehnTwists} that represent the added 
generators.\footnote{The added generators are internal 
transformations (i.e. have support within the prime factors) 
and hence behave naturally under exchanges. They commute with all 
other internal diffeomorphisms and slides \emph{of} the prime in 
question, since their supports may be chosen to be disjoint from 
the diffeomorphisms representing the other internal diffeomorphisms
and slides. They also commute with slides of other primes \emph{through} 
the one in question, since their action on such slides (by conjugation) 
is a slide along a curve isotopic to the original one, which defines 
the same mapping class. Finally, the conjugation of a handle's 
`twist' (right picture in Fig.\,\ref{fig:DehnTwists}) with a spin 
of that handle is isotopic to the original twist.}

The procedure just outlined reduces the problem of finding a 
presentation of $\MCGF$ to that of finding presentations 
$\mathcal{G}_F(\Pi_i)$ for each irreducible prime. As already 
stated, they are explicitly known for all spherical space forms. 
We also mentioned that for Haken manifolds $\mathcal{G}_F(\Pi_i)$ 
is a $\mathbb{Z}_2$ extension of $\mathrm{Aut}^+(\pi_1)$, which in 
simple cases allows to find an explicit presentation. For example, 
for the 3-torus we have 
$\mathrm{Aut}^+(\pi_1=\mathbb{Z}^3)=SL(3{,}\mathbb{Z})$ and 
the appropriate central $\mathbb{Z}_2$ extension can be shown to be 
given by the Steinberg group $St(3{,}\mathbb{Z})$, a presentation 
of which may e.g. be found in \S\,10 of \cite{Milnor:KTheory}. 

\begin{figure}[ht]
\centering\epsfig{figure=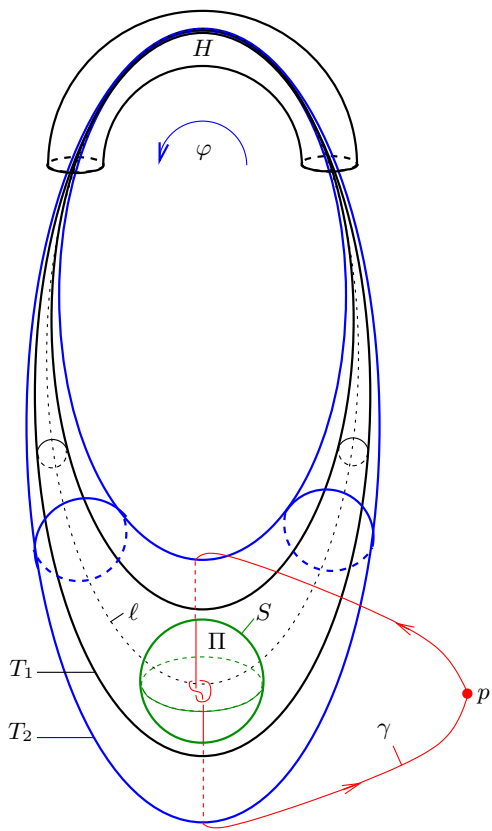,width=
0.48\linewidth}
\put(-176,59){\small $T_1$}
\put(-176.5,35){\small $T_2$}
\put(-83,80){\small $S$}
\put(-101,69){\small $\Pi$}
\put(-105.5,256){\small $\varphi$}
\put(-107,293.5){\small $H$}
\put(1,50){\small $p$}
\put(-37,37){\small $\gamma$}
\put(-131,80){\small $\ell$}
\caption{\label{fig:Slide}%
The slide of a prime as seen form the `inside view' (compare 
the left picture in Fig.\,\ref{fig:ConnectedSum}) through 
a handle $H$ in the background. The prime to be slid, $\Pi$, 
hides behind its connecting 2-sphere $S$. The loop $\ell$, 
representing the generator of $\pi_1(H)$ along which the prime is 
slid, is thickened to two coaxial tori, $T_1$ and $T_2$, such that 
the prime is contained in the inner torus $T_1$. 
The diffeomorphism in the toroidal region is given by 
(\ref{eq:DefSlide}), where the angle $\varphi$ measures the 
axial direction.  The slide acts on $[\gamma]\in\pi_1(\Pi,p)\subset
\pi_1(\Sigma,p)$ (the non-triviality of which being indicated by 
the little knot inside $\Pi$) by conjugation with $[\ell]$, when 
the latter is appropriately considered as element of 
$\pi_1(\Sigma,p)$.}
\end{figure}

All the generators listed in 2.-5. can be realized by appropriate
diffeomorphisms. This is not difficult to see for 2. and 3., as 
diffeomorphisms that `spin' a handle or `exchange' two diffeomorphic 
primes are easily visualized. A visualization of the slide 
transformations 4. and 5. is attempted in Fig.\,\ref{fig:Slide}: 
The general idea---and this is where the name `slide' derives from---is 
similar to that of rotation of parallel (i.e. concentric) spheres
explained in Fig.\,\ref{fig:DehnTwists}. But now we take two 
`parallel' (i.e. coaxial) tori, $T_1$ and $T_2$, and consider a 
diffeomorphism whose support is confined to the region between 
them. This toroidal region is of topology 
$[1,2]\times S^1\times S^1$ (i.e. does not contain prime summands) 
and hence is foliated by a one parameter ($r$) family of parallel 
(coaxial)  tori $T_r=r\times S^1\times S^1$, where $r\in[1,2]$. 
Each of these tori we think of as being coordinatized in the standard
fashion by two angles, $\theta$ and $\varphi$ with range $[0,2\pi]$ 
each, where $\theta$ labels the latitude and $\varphi$ the longitude 
(the circles of constant $\varphi$ are the small ones that become 
contractible in the solid torus). The slide now corresponds to a 
diffeomorphism which is the identity outside the toroidal region and 
which inside ($1\leq r\leq 2$) is given by 
\begin{equation}
\label{eq:DefSlide}
\bigl(r,\theta,\varphi\bigr)\mapsto 
\bigl(r,\theta,\varphi+\beta(2-r)\bigr)\,,
\end{equation}
where $\beta$ is a $C^{\infty}$ step-function 
$\beta:[0,1]\rightarrow[0,2\pi]$ with $\beta(0)=0$ and $\beta(1)=2\pi$. 
In Fig.\,\ref{fig:Slide}, the loop $\gamma$ generates a non-trivial 
element $[\gamma]\in\pi_1(\Pi,p)$ (the non-triviality is indicated 
by the little knot inside $\Pi$ for lack of better representation). 
This loop $\gamma$, after having been acted on by the slide, will 
first follow $\ell$ and go through the handle, then travel 
trough $\Pi$ as before, and finally travel the handle in a reversed 
sense. That is, the slide conjugates $[\gamma]$ with $[\ell]$. 

The slides of irreducible primes described in 4.), or the slides of
ends of handles described in 5.), are then obtained by choosing the 
tori such that the only connecting sphere contained inside the inner 
torus is that of the prime, or handle-end, to be slid. The common 
axis of the tori trace out a non-contractible loop through another 
prime, the prime through which the first one is slid. All this is 
depicted in Fig.\,\ref{fig:Slide}.

Of crucial importance is the different behavior of slides in 4.) on one 
hand, and slides in 5.) on the other. Algebraically this has to do with 
the different behavior, as regards the automorphism group of the
free product, of the free factors $\mathbb{Z}$ on one hand and the 
non-free factors $G_{(i)}$ on the other. Whereas left or right 
multiplication of all elements in one $\mathbb{Z}$ factor with any 
element from the complementary free product defines an automorphism of $G$, 
this is not true for the non-free factors. Here only conjugation 
defines an automorphism. Geometrically this means that we have to 
consider  slides of both ends of a handle separately in order to 
be able to generate the automorphism group. It is for this reason 
that we pictured the handles in Fig.\,\ref{fig:ConnectedSum} as being 
attached to the base manifold with two rather than just one 
connecting sphere. 

In passing we remark that certain important interior diffeomorphism 
(i.e. falling under case\,1.) above) have an interpretation in terms 
of certain \emph{internal} slides. Imagine Fig.\,\ref{fig:Slide} as 
an \emph{inside} view from some prime $\Pi$, i.e. everything seen on 
Fig.\,\ref{fig:Slide} is inside $\Pi$. The handle $H$, too, is now 
interpreted as some topological structure of $\Pi$ itself that gives 
rise to non-contractible loops within $\Pi$. $S$ is again the 
connecting sphere of $\Pi$, now seen from the inside, beyond which 
the part of the manifold $\Sigma$ outside $\Pi$ lies. The diffeomorphism 
represented by Fig.\,\ref{fig:Slide} then slides the connecting 
sphere $S$ once around the loop $\ell$ in $\Pi$. Its effect is to 
conjugate each element of $\pi_1(\Pi)$ by $[\ell]\in\pi_1(\Pi)$. 
Hence we see that such internal slides generate all \emph{internal} 
automorphisms for each irreducible factor $\Pi$. In case of handles, 
there are no non-trivial inner automorphisms, and the only non-trivial 
outer automorphism $(\mathbb{Z}\mapsto -\mathbb{Z})$ is realized by 
spinning the handle, as already mentioned.

Having said that, we will from now on always understand by `slides' 
\emph{external} transformations as depicted in \ref{fig:Slide}, unless 
explicitly stated otherwise (cf. last remark at the end of this section).  
But let us for the moment forget about slides altogether and focus 
attention only on those mapping classes listed in 1.)-3.), i.e. 
internal transformations and exchanges. In doing this we think of 
a spin of a handle as internal, which we may do as long as 
no slides are considered. It is tempting to think of the manifold 
$\Sigma$ as being composed of $N=n+m$ `particles' from $d$ species, 
each with its own characteristic internal symmetry group 
$G_r$, $1\leq r\leq d$. In this analogy diffeomorphic primes 
correspond to particles of one species and the symmetry groups $G_r$ 
correspond to $\mathcal{G}_{\mathrm{F}}(P_r)$. Let there be $n_r$ 
primes in the $r$-th diffeomorphism class, so that $\sum_{r=1}^dn_r=N$. 
In this `particle picture' the symmetry group would be a semi-direct 
product of the internal symmetry group, $G^I$, with an external 
symmetry group, $G^E$, both respectively given by 
\begin{subequations}
\begin{alignat}{2}
\label{eq:DefIntSymm}
& G^I &&\,:=\,
\prod^{n_1}G_1\times\cdots\times\prod^{n_d}G_d\,,\\
\label{eq:DefExtSymm}
& G^E &&\,:=\,
S_{n_1}\times\cdots\times S_{n_d}\,,
\end{alignat}
\end{subequations}
where here $S_n$ denotes the order $n!$ permutation group of $n$ 
objects. The semi-direct product is characterized through the 
homomorphism $\theta: G^E\rightarrow\mathrm{Aut}\bigl(G^I\bigr)$, 
where $\theta=\theta_1\times\cdots\times\theta_d$ and 
\begin{equation}
\label{eq:AutHom}
\begin{split}
\theta_i:S_{n_i}&\rightarrow\mathrm{Aut}\left(\prod^{n_i}G_i\right)\\
\sigma&\mapsto\theta_i(\sigma):
\bigl(g_1,\cdots,g_{n_i}\bigr)\mapsto
\bigl(g_{\sigma(1)},\cdots,g_{\sigma(n_i)}\bigr)\,.\\
\end{split}
\end{equation}
The semi-direct product $G^I\rtimes G^E$ with respect to 
$\theta$ is now defined by the following multiplication law:
let $\gamma_i\in\prod^{n_i}G_i$, $i=1,\cdots,d$, then
\begin{equation}
\label{eq:SemiDirExtInt}
\begin{split}
&\bigl(\gamma'_1,\cdots,\gamma'_d\,;\,
\sigma'_1,\cdots,\sigma'_d\bigr) 
\bigl(\gamma_1,\cdots,\gamma_d\,;\,
\sigma_1,\cdots,\sigma_d\bigr)\\
&=\bigl(\gamma'_1\,\theta_1(\sigma'_1)(\gamma_1),\cdots,
 \gamma'_d\,\theta_d(\sigma'_d)(\gamma_d)\,;\,
\sigma'_1\sigma_1,\cdots,\sigma'_d\sigma_d\bigr)\,.\\
\end{split}
\end{equation}    
We call $G^P=G^I\rtimes G^E$ the \emph{particle group}. From the 
discussion above it is clear that this group is a subgroup of the 
mapping class group. But we also had to consider slides which 
were neither internal nor exchange diffeomorphisms and which are 
not compatible with this simple particle picture, since they mix 
internal and external points of the manifold. How much do the slides 
upset the particle picture? For example, consider the normal 
closure, $G^S$, of slides (i.e. the smallest normal subgroup in 
the group of mapping classes that contains all slides). 
Does it have a non-trivial intersection with $G^P$, i.e. is
$G^P\cap G^S\ne\{1\}$? If this is the case $\MCGF/G^S$ will be a 
non-trivial factor of $G^P$. Representations whose kernels contain 
$G^S$ will then not be able to display all particle symmetries. 
This would only be the case if $G^P\cap G^S=\{1\}$. 

Questions of this type have been addressed and partly answered in 
\cite{Giulini:1997a} (see also \cite{SorkinSurya:1998}). Here are 
some typical results: 
\begin{prop}
\label{prop:NoHandles}
$G^P\cap G^S=\{1\}$ and  $\MCGF=G^S\rtimes G^P$ if 
$\Sigma$ contains no handle in its prime decomposition. 
\end{prop} 
\begin{prop}
\label{prop:ThreeOrMoreHandles}
$G^S$ is perfect if $\Sigma$ contains at least 3 handles
in its prime decomposition. 
\end{prop} 
\noindent
The last proposition implies that slides cannot be seen in 
abelian representations of mapping class groups of manifolds 
with at least three handles. In \cite{Giulini:1997a} an explicit 
presentation with four generators of the mapping class 
group of the connected sum of $n\geq 3$ handles was given and 
the following result was shown: 
\begin{prop}
\label{prop:NoSlidesRep}
Let $\Sigma=\stackrel{n}{\uplus}S^1\times S^2$ where $n\geq 3$. 
Then $\MCGF/G^S\cong\mathbb{Z}_2\times\mathbb{Z}_2$ where one 
$\mathbb{Z}_2$ is generated by the twist (right picture in 
Fig.\,\ref{fig:2PiTwist}) of, say, the first handle and the 
other $\mathbb{Z}_2$ by either the exchange of, say, the first 
and second handle or the spin of, say, the first handle. Hence 
we have a strict correlation between spins and exchanges of handles. 
Generally, given a representation $\rho$ of $\MCGF$, the following 
statements are equivalent:  
\begin{itemize}
\item[a.)] 
$\rho$ is abelian, 
\item[b.)]
slides are in the kernel of $\rho$,
\item[c.)]
$\rho$ strictly correlates exchanges and spins
(i.e. $\rho(\mathrm{spin})=\rho(\mathrm{exchange})$),
\item[d.)]
slides and exchanges commute under $\rho$. 
\end{itemize}
\end{prop}

Reference \cite{Giulini:1997a} also deals with connected sums of 
arbitrarily many real projective spaces $\RP^3$. A presentation in 
terms of three generators was written down and various features 
studied. Since projective spaces are the most simple lens spaces 
($\RP^3=L(2{,}1)$) they are not spinorial, as one easily visualizes. 
Since the automorphism group of $\pi_1(\RP^3)=\mathbb{Z}_2$ is 
trivial there are no non-trivial mapping classes from internal 
diffeomorphisms ($\RP^3$ satisfies the homotopy-implies-isotopy 
property). Therefore, the particle group $G^P$ is just the 
permutation group $S_n$ and the mapping class group is the 
semi-direct product $G^S\rtimes S_n$, according to 
Prop.\,\ref{prop:NoHandles}. A a very interesting systematic study 
of representaions of this group was started in \cite{SorkinSurya:1998}
using Mackey theory (theory of induced representations). 
For illustrative purposes we consider in some detail the case of 
the connected sum of just two projective spaces in the next section,
also considered in \cite{Aneziris-etal:1989a} and 
\cite{SorkinSurya:1998}.

Even though we here restrict attention to frame-fixing diffeomorphisms,
which is physically well motivated, we nevertheless wish to end this 
section with a few remarks that give an idea of the essential changes 
that result if we relaxed from $\DiffF$ (or $\DiffInf$) 
to $\Diff$. In the frame-fixing context, non-trivial mapping classes 
of prime factors (i.e. generated by internal diffeomorphisms) are 
non-trivial mapping classes in the total manifold $\Sigma$. In other 
words, there is an injection 
$\mathcal{G}_\mathrm{F}(P)\rightarrow\MCGF$~\cite{HendriksMcCullough:1987}. 
As already remarked above, this is not true if $\MCGF$ is replaced by 
$\MCG$. In fact, it follows from the above that any diffeomorphism 
in $\DiffF$ whose image under $h_\mathrm{F}$ in 
$\mathrm{Aut}(\pi_1(\Sigma))$ is an \emph{inner} automorphism is 
isotopic in $\Diff$ to transformations of the type depicted in 
Fig.\,\ref{fig:DehnTwists}. Hence there are generally many inner 
diffeomorphisms of prime factors $P$ which represent non-trivial 
elements of $\mathrm{Diff}_\mathrm{F}(P)$ but trivial elements 
in $\Diff$. Also, the distinction between inner and non-inner 
(exchanges and slides) diffeomorphisms ceases to be meaningful in 
$\Diff$. A trivial example is the diffeomorphism depicted in 
Fig.\,\ref{fig:2PiTwist}, i.e. the $2\pi$-rotation parallel to the
common connecting sphere of two irreducible prime manifolds. 
Up to isotopy in $\Diff$ it can clearly be considered as inner 
diffeomorphism of \emph{either} prime. A less trivial example is the 
following: consider $\Sigma=\Pi_1\uplus\Pi_2$, where 
$\{g_{(1)1},\cdots,g_{(1)n_1}\}$ and $\{g_{(2)1},\cdots,g_{(2)n_2}\}$
are the generators of $\pi_1(\Pi_1)$ and $\pi_1(\Pi_2)$ 
respectively. As explained above, a slide of $\Pi_1$ through $\Pi_2$ 
along, say, $g^{-1}_{(2)1}$ acts on these sets of generators by 
conjugating each in the first set with $g^{-1}_{(2)1}$. On the 
other hand, each inner automorphisms of $\pi_1(\Sigma)$ 
can be produced by a diffeomorphism that is isotopic to the 
identity in $\Diff$.\footnote{That there is a diffeomorphism 
generating any inner automorphism of $\pi_1(\Sigma)$ is shown 
exactly as in our discussion of `internal slides' above. See
Fig.\,\ref{fig:Slide}, where now everything is in $\Sigma$ and 
the inside of $T_1$ is taken to be a solid torus, i.e. there is 
\emph{no} prime factor $\Pi$ inside $T_1$. Then it is obvious that 
the slide depicted is isotopic to the identity in $\Diff$ by 
a diffeomorphism whose support is inside $T_2$ but extends into 
the inside of $T_1$: just take the isotopy 
$[0,1]\ni s\mapsto\phi_s:(r,\theta,\varphi)\mapsto
\bigl(r,\theta,s\beta(2-r)\bigr)$, where 
$\beta:[0,1]\rightarrow[0,2\pi]$ is the step function as in 
(\ref{eq:DefSlide}), continued by the constant value $2\pi$ 
to $[1,2]$.} Taking for that the diffeomorphism that 
conjugates $\pi_1(\Sigma)$ by $g_{(2)1}$ we see that the slide just 
considered is isotopic (in $\Diff$, not in $\DiffF$ of $\DiffInf$) 
to a diffeomorphism that leaves the $g_{(1)i}$ untouched and 
conjugates the $g_{(2)i}$ by $g_{(2)1}$. This, in turn, can be 
represented by an inner diffeomorphisms of $\Pi_2$ (an internal 
slide, possibly with a $2\pi$-rotation). This shows that the original 
slide of $\Pi_1$ trough $\Pi_2$ is isotopic in $\Diff$ (not in 
$\DiffF$ or $\DiffInf$) to an internal diffeomorphism of 
$\Pi_2$.\footnote{This point is not correctly taken care of 
in \cite{DowkerSorkin:1998}, where it is argued that a slide 
of one prime through the other is \emph{always} either isotopic 
to the identity or the relative $2\pi$-rotation of both primes 
(\cite{DowkerSorkin:1998}, p.\,1162). But this is only true if 
the loop along which is slid generates a central element of 
$\pi_1(\Pi_2)$. In particular, the isotopy claimed in the last 
sentence of the footnote on p.\,1162 of \cite{DowkerSorkin:1998} 
cannot exist in general. However, the application in 
\cite{DowkerSorkin:1998} is eventually restricted to primes with 
abelian fundamental group so that this difficulty does not affect 
the specific conclusions drawn in \cite{DowkerSorkin:1998}. 
I thank Fay Dowker and Bob Gompf for discussions of that point.} 
For example, if $\Pi_2$ is a lens space, which is non-spinorial 
and has abelian fundamental group, the internal diffeomorphism 
just considered is clearly isotopic to the identity. Hence the 
original slide of the first prime through the lens space is 
isotopic to the identity in $\Diff$, but not in $\DiffInf$ or 
$\DiffF$. All this shows that the division of mapping-class 
generators into internal diffeomorphisms, exchanges, and 
slides only makes sense in $\DiffInf$ and $\DiffF$, but not in 
$\Diff$.

\section{A simple yet non-trivial example} 
\label{sec:ExampleTwoRP3}
In this section we wish to discuss in detail the mapping 
class group $\MCGF$ for $\Sigma=\RP^3\uplus\RP^3$. 
Before doing this, let us say a few words on how the single 
$\RP^3$ manifold can arise in an exact black-hole solution 
in General Relativity. 

\subsection{The $\RP^3$ geon}
\label{sec:RP3Geon}
\begin{figure}[ht]
\begin{minipage}[c]{0.48\linewidth}
\centering\epsfig{figure=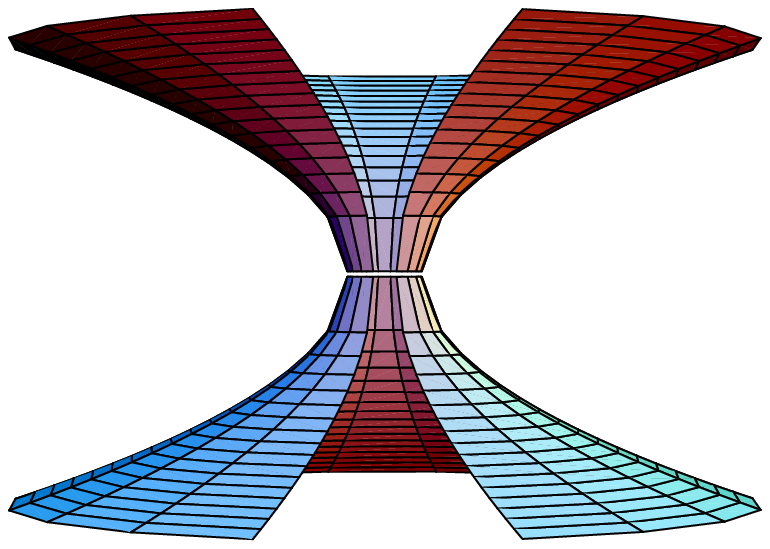,width=1.0\linewidth}
\end{minipage}
\hfill
\begin{minipage}[c]{0.5\linewidth}
\centering\epsfig{figure=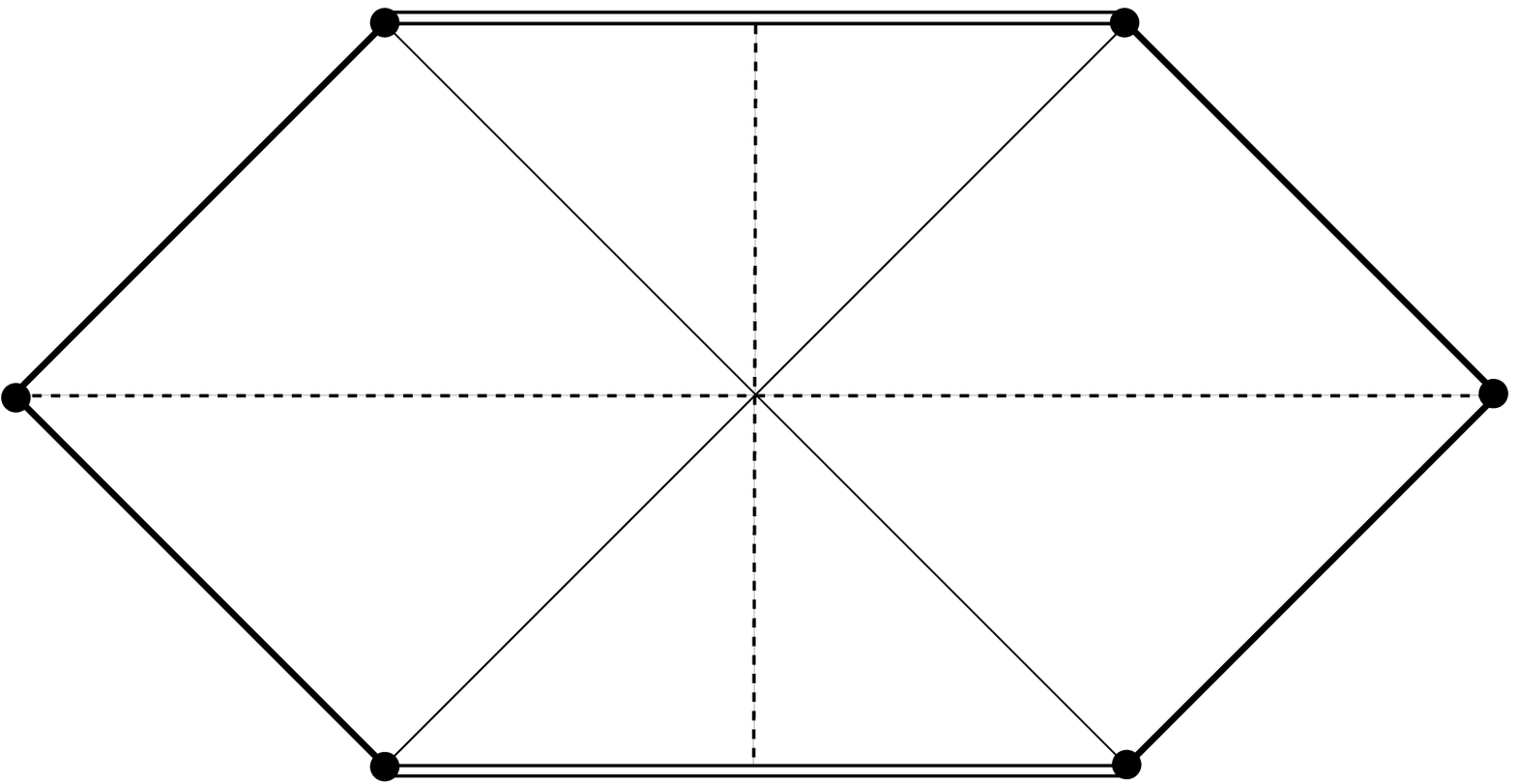,width=1.0\linewidth}
\end{minipage}
\put(-143,4){\tiny $T=0$}
\put(-90.5,18){\tiny\begin{rotate}{90}$X=0$\end{rotate}}
\put(1,2){\tiny $i_0$}
\put(-185,2){\tiny $i_0$}
\put(-47,51){\tiny $i_+$}
\put(-134,51){\tiny $i_+$}
\put(-47,-48){\tiny $i_-$}
\put(-134,-48){\tiny $i_-$}
\put(-20,23){\tiny $I^+$}
\put(-20,-21){\tiny $I^-$}
\put(-164,23){\tiny $I^+$}
\put(-166,-21){\tiny $I^-$}
\put(-48,20){$\mathrm{I}$}
\put(-84,30){$\mathrm{II}$}
\put(-138,20){$\mathrm{III}$}
\put(-86,-30){$\mathrm{IV}$}
\put(-290,45){$T=0$}
\put(-325,1){\tiny $X=0\quad\rightarrow$}
\caption{\label{fig:Kruskal}%
To the right is the conformal (Penrose) diagram of Kruskal spacetime 
in which each point of this 2-dimensional representation corresponds 
to a 2-sphere (an orbit of the symmetry group of spatial rotations). 
The asymptotic regions are $i_0$ (spacelike infinity), $I^{\pm}$ 
(future/past lightlike infinity), and $i^{\pm}$ (future/past timelike 
infinity). The diamond and triangular shaped regions $\mathrm{I}$
and $\mathrm{II}$ correspond to the exterior ($r>2m$) and interior 
($0<r<2m$) Schwarzschild spacetime respectively, the interior being 
the black hole. The triangular region $\mathrm{IV}$ is the time reverse of 
$\mathrm{II}$, a white hole. Region $\mathrm{III}$ is another 
asymptotically flat end isometric to the exterior Schwarzschild region
$\mathrm{I}$. The double horizontal lines on top an bottom represent the 
singularities ($r=0$) of the black and white hole respectively. 
The left picture shows an embedding diagram of the hypersurface $T=0$ 
(central horizontal line in the conformal diagram) that serves to 
visualize its geometry. Its minimal 2-sphere at the throat corresponds 
to the intersection of the hyperplanes $T=0$ and $X=0$ (bifurcate 
Killing Horizon).}
\end{figure}

Recall that we limited attention to asymptotically flat manifolds 
with a single end (no `internal infinity'). Is this not too severe 
a restriction? After all, we know that the (maximally extended)
manifold with one (uncharged, non-rotating) black hole is the 
Kruskal manifold~\cite{Kruskal:1960} (see also Chapter\,5.5. in 
\cite{HawkingEllis:TLSSOS}), in which space has \emph{two} ends. 
Figure\,\ref{fig:Kruskal} shows the familiar conformal diagram, 
where the asymptotically flat ends lie in regions~$\mathrm{I}$ and 
$\mathrm{III}$. In Kruskal coordinates\footnote{Kruskal~\cite{Kruskal:1960} 
uses $(v,u)$ Hawking Ellis \cite{HawkingEllis:TLSSOS} $(t',x')$
for what we call $(T,X)$.} $(T,X,\theta,\varphi)$, where $T$ and $X$ 
each range in $(-\infty,\infty)$ obeying $T^2-X^2<1$, the Kruskal 
metric reads (as usual, we write 
$d\Omega^2$ for $d\theta^2+\sin^2\theta\,d\varphi^2$): 
\begin{equation}
\label{eq:KruskalMetric1}
g=\frac{32 m^2}{r}\,\exp(-r/2m)\,\bigl(-dT^2+dX^2\bigr)+r^2d\Omega^2\,,
\end{equation}
where $r$ is a function of $T$ and $X$, implicitly 
defined by 
\begin{equation}
\label{eq:KruskalMetric2}
\bigl((r/2m)-1\bigr)\,\exp(r/2m)=X^2-T^2\,
\end{equation}
and where $m>0$ represents the mass of the hole in geometric units.
The metric is spherically symmetric and allows for the additional 
Killing field\footnote{That $K$ is Killing is immediate, 
since $r$ depends only on the combination $X^2-T^2$ which is 
annihilated by $K$.} 
\begin{equation}
\label{eq:KruskalKilling}
K=\tfrac{1}{4m} \bigl(X\partial_T+T\partial_X\bigr)\,,
\end{equation}
which is timelike for $\vert X\vert>\vert T\vert$ and spacelike for 
$\vert X\vert<\vert T\vert$.
\bigskip
 
The familiar exterior Schwarzschild solution is given by 
region~$\mathrm{I}$ in Fig.\,\ref{fig:Kruskal}, where the 
transformation from Schwarzschild coordinates $(t,r,\theta,\varphi)$,
where $2m<r<\infty$ and $-\infty<t<\infty$, to Kruskal coordinates is given by  
\begin{subequations}
\begin{alignat}{2}
\label{eq:Kruskal1}
& T&&\,=\,\sqrt{(r/2m)-1}\,\exp(r/4m)\,\sinh(t/4m)\,,\\
\label{eq:Kruskal2}
& X&&\,=\,\sqrt{(r/2m)-1}\,\exp(r/4m)\,\cosh(t/4m)\,.
\end{alignat}
\end{subequations}
This obviously just covers region~$\mathrm{I}$: 
$X>\vert T\vert$. In Schwarzschild coordinates the Killing 
field (\ref{eq:KruskalKilling}) just becomes $K=\partial_t$. 

Now consider the following discrete isometry of the Kruskal manifold: 
\begin{equation}
\label{eq:KrskalZ2Isometry}
J:(T,X,\theta,\varphi)\mapsto (T,-X,\pi-\theta,\varphi+\pi)\,.
\end{equation}
It generates a freely acting group $\mathbb{Z}_2$ of smooth 
isometries which preserve space- as well as time-orientation. 
Hence the quotient is a smooth space- and time-orientable manifold, 
the $\RP^3$\emph{geon}.\footnote{The $\RP^3$ geon is different 
from the two mutually different `elliptic interpretations' of the 
Kruskal spacetime discussed in the literature by Rindler, Gibbons, 
and others. In \cite{Rindler:1965} the identification map considered 
is $J':(T,X,\theta,\varphi)\mapsto(-T,-X,\theta,\varphi)$, which 
gives rise to singularities on the set of fixed-points 
(a two-sphere) $T=X=0$. Gibbons \cite{Gibbons:1986} takes  
$J'':(T,X,\theta,\varphi)\mapsto(-T,-X,\pi-\theta,\varphi+\pi)$,
which is fixed-point free, preserves the Killing field 
(\ref{eq:KruskalKilling}) (which our map $J$ does not), but does 
not preserve time-orientation. $J''$ was already considered in 
1957 by Misner \& Wheeler (Section\,4.2 in \cite{MisnerWheeler:1957}), 
albeit in so-called `isotropic Schwarzschild coordinates', which 
only cover the exterior regions $\mathrm{I}$ and $\mathrm{III}$ 
of the Kruskal manifold.}
Its conformal diagram is just given by 
cutting away the $X<0$ part (everything to the left of the vertical 
$X=0$ line) in Fig.\,\ref{fig:Kruskal} and taking into account that 
each point on the remaining edge, $X=0$, now corresponds to a 2-sphere
with antipodal identification, i.e. a $\RP^2$ which is non-orientable. 
The spacelike hypersurface $T=0$ has now the topology of the 
once punctured $\RP^3$. In the left picture of Fig.\,\ref{fig:Kruskal}
this corresponds to cutting away the lower half and eliminating the 
inner boundary 2-sphere $X=0$ by identifying antipodal points. 
The latter then becomes a minimal one-sided non-orientable surface 
in the orientable space-section of topology $\RP^3-\{\mathrm{point}\}$.
The $\RP^3$ geon isometrically contains the exterior Schwarzschild 
spacetime (region\,$\mathrm{I}$) with timelike Killing field $K$.
But $K$ ceases to exits globally on the geon spacetime since it 
reverses direction under (\ref{eq:KrskalZ2Isometry}). 

Even though the Kruskal spacetime and its quotient are, geometrically
speaking, locally indistinguishable, their physical properties are 
different. In particular, the thermodynamic properties of quantum 
fields are different. For details we refer to \cite{Louko:2000} 
and references therein. We also remark that the mapping-class group 
$\mathcal{G}_{\mathrm{F}}(\RP^3)$ is trivial~\cite{Witt:1986b},
as are the higher homotopy groups of 
$\SuperRes(\RP^3)$~\cite{Giulini:1995a}.
Equation (\ref{eq:HomotopyGroups}) then shows that the 
configuration space $\SuperRes(\RP^3)$ is (weakly) homotopically 
contractible. In fact, the three-sphere $S^3$ and the real 
projective 3-space $\RP^3$ are the only 3-manifolds for which 
this is true; see~\cite{Giulini:1994a} (table on p.\,922).

\subsection{The connected sum $\RP^3\uplus\RP^3$}
\label{sec:TwoGeons}
Asymptotically flat initial data on the once punctured manifold 
$\RP^3\uplus\RP^3-\{\mathrm{point}\}$ can be explicitly constructed.
For this one considers time-symmetric conformally-flat initial 
data. The constraints (\ref{eq:ConstEq}) then simply reduce to 
the Laplace equation for a positive function $\Phi$, where 
$\Phi^4$ is the conformal factor. The `method of images' known 
from electrostatics can then be employed to construct special 
solutions with reflection symmetries about two 2-spheres. The topology 
of the initial data surface is that of  $\mathbb{R}^3$ with two 
disjoint open 3-discs excised. This excision leaves two inner 
boundaries of $S^2$ topology on each of which antipodal  points are 
identified. The metric is
constructed in such a way that it projects in a smooth fashion to 
the resulting quotient manifold whose topology is that of 
$\RP^3\uplus\RP^3-\{\mathrm{point}\}$. Details and analytic 
expressions are given in~\cite{Giulini:1998}. These data describe 
two black holes momentarily at rest. The spacetime they involve into 
(via (\ref{eq:EvolEq})) is not known analytically, but since the 
analytical form of the data is very similar indeed to the form of 
the Misner-wormehole data (cf.~\cite{Giulini:1998}), which were often 
employed in numerical studies, I would not expect the numerical 
evolution to pose any additional difficulties. 
All this is just to show that the once punctured manifold 
$\RP^3\uplus\RP^3$ is not as far fetched in General Relativity as 
it might seem at first: it is as good, and no more complicated than, 
the Misner wormehole which is the standard black-hole data set in
numerical studies of head-on collisions of equal-mass black 
holes.\footnote{The $\RP^3$ data even have certain advantages: they
generalize to data where the masses of the black holes are not equal
(for the wormehole identification the masses need to be equal) and 
even to data for any number of holes with arbitrary masses (in which 
case the holes may not be `too close').} 

\begin{figure}[ht]
\centering\epsfig{figure=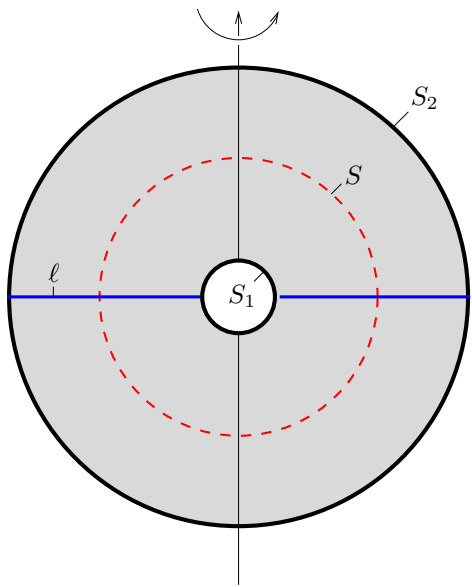,width=0.5\linewidth}
\put(-49,152){$S$}
\put(-93,107){$S_1$}
\put(-24,182){$S_2$}
\put(-161,115){$\ell$}
\caption{\label{fig:TwoRP3}%
Visualization of $\RP^3\uplus\RP^3$, which corresponds to the shaded 
region between the 2-spheres $S_1$ and $S_2$, on each of which antipodal 
points are identified. As indicated, the whole picture is to be thought 
of as rotating about the vertical axis, except for the solid vertical 
line $\ell$, which, like any other radial line, corresponds to a closed 
loop, showing that $\RP^3\uplus\RP^3$ is fibered by circles over $\RP^2$. 
Each fiber intersects the connecting 2-sphere, $S$, in two distinct 
points.}
\end{figure}

We wish to study and visualize the mapping class group 
$\mathcal{G}_{\mathrm{F}}(\Sigma)$ where $\Sigma=\RP^3\uplus\RP^3$. 
We represent $\Sigma$ by the annular region depicted in 
Fig.\,\ref{fig:TwoRP3}, which one should think of as 
representing a 3-dimensional spherical shell inbetween the outer 
boundary 2-sphere $S_2$ and the inner boundary 2-sphere $S_1$. 
In addition, on each boundary 2-sphere we identify antipodal 
points. The result is the connected sum $\RP^3\uplus\RP^3$ where 
we might take $S$ for the connecting 2-sphere that lies `half way'
between $S_1$ and $S_2$ in Fig.\,\ref{fig:TwoRP3}. The radial lines,
like $\ell$, fiber the space in loops showing that $\RP^3\uplus\RP^3$ 
is an $S^1$ bundle over $\RP^2$.\footnote{$\RP^3\uplus\RP^3$ is an 
example of a Seifert fibered space without exceptional fibers, that 
was already mentioned explicitly in Seifert's thesis~\cite{Seifert:1932}.}
Interestingly, it is doubly covered by the prime manifold $S^1\times S^2$, 
the corresponding deck transformation of the latter being 
$(\psi,\theta,\varphi)\mapsto (2\pi-\psi,\pi-\theta,\varphi+\pi)$,
where $\psi\in[0,2\pi]$ coordinatizes $S^1$ and $(\theta,\varphi)$ 
are the standard spherical polar coordinates on $S^2$. 
Note that this deck transformation does not commute with the 
$SO(2)$ part of the obvious transitive $SO(2)\times SO(3)$ action 
on $S^1\times S^2$, so that only a residual $SO(3)$ action remains 
on $\RP^3\uplus\RP^3$ whose orbits are 2-spheres except for two 
$\RP^2$s.\footnote{$\RP^3$ is trivially homogeneous, being 
$SO(3)$. The punctured space $\RP^3-\{\mathrm{point}\}$ is also 
homogeneous, since it may be identified with the space of all 
hyperplanes (not necessarily through the origin) in $\mathbb{R}^3$, 
on which the group $E_3=\mathbb{R}^3\rtimes SO(3)$ of Euclidean motions 
clearly acts transitively with stabilizers isomorphic to $E_2$; hence 
$\RP^3-\{\mathrm{point}\}\cong E_3/E_2$.} It is known to be the only 
example of a (closed orientable) 3-manifold that is a proper connected 
sum and covered by a prime.\footnote {Clearly, no proper 
connected sum ($\pi_2\ne 0$) can be covered by an irreducible 
prime ($\pi_2=0$).}

\begin{figure}[htb]
\begin{minipage}[c]{0.40\linewidth}
\centering\epsfig{figure=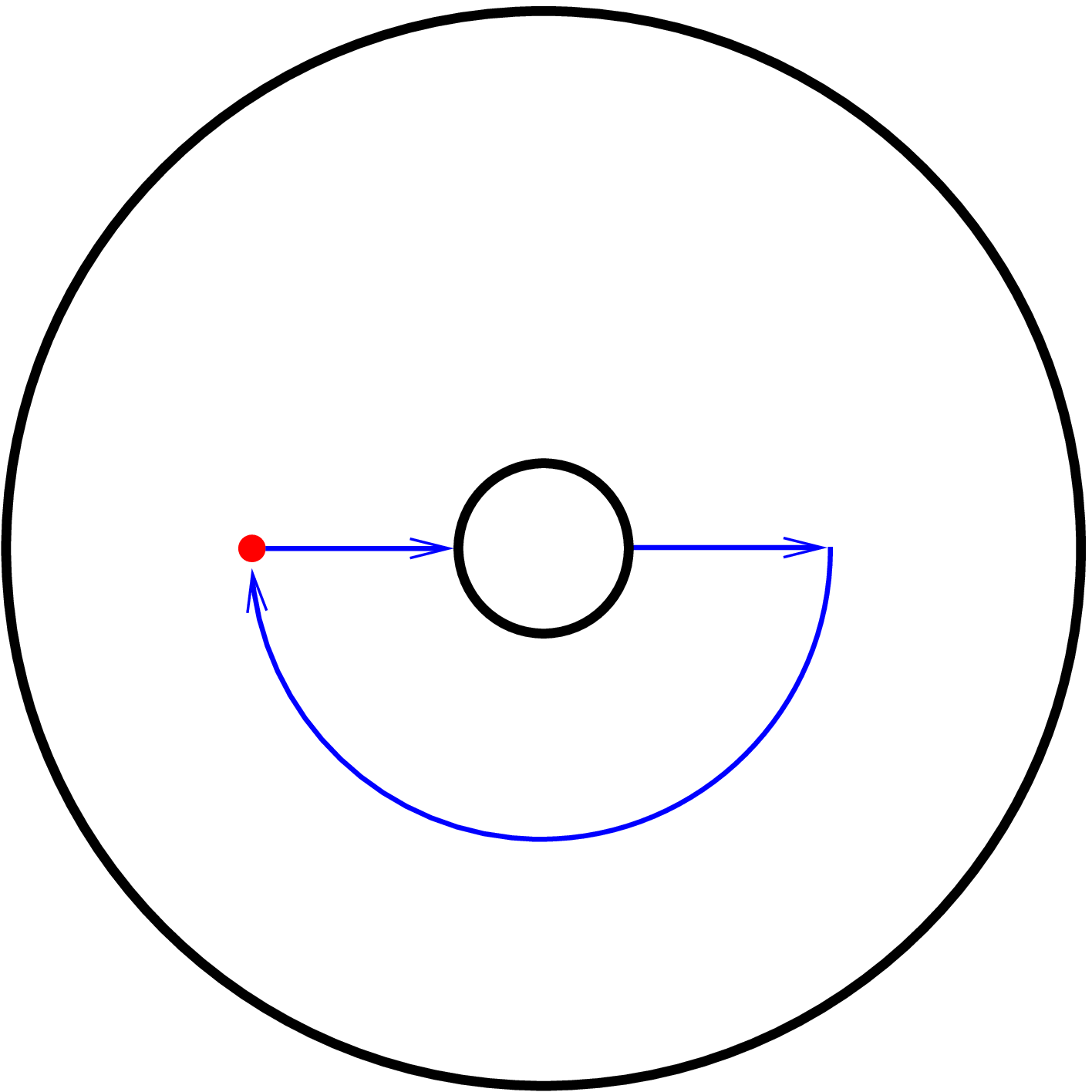,width=0.9\linewidth}
\end{minipage}
\hfill
\begin{minipage}[c]{0.40\linewidth}
\centering\epsfig{figure=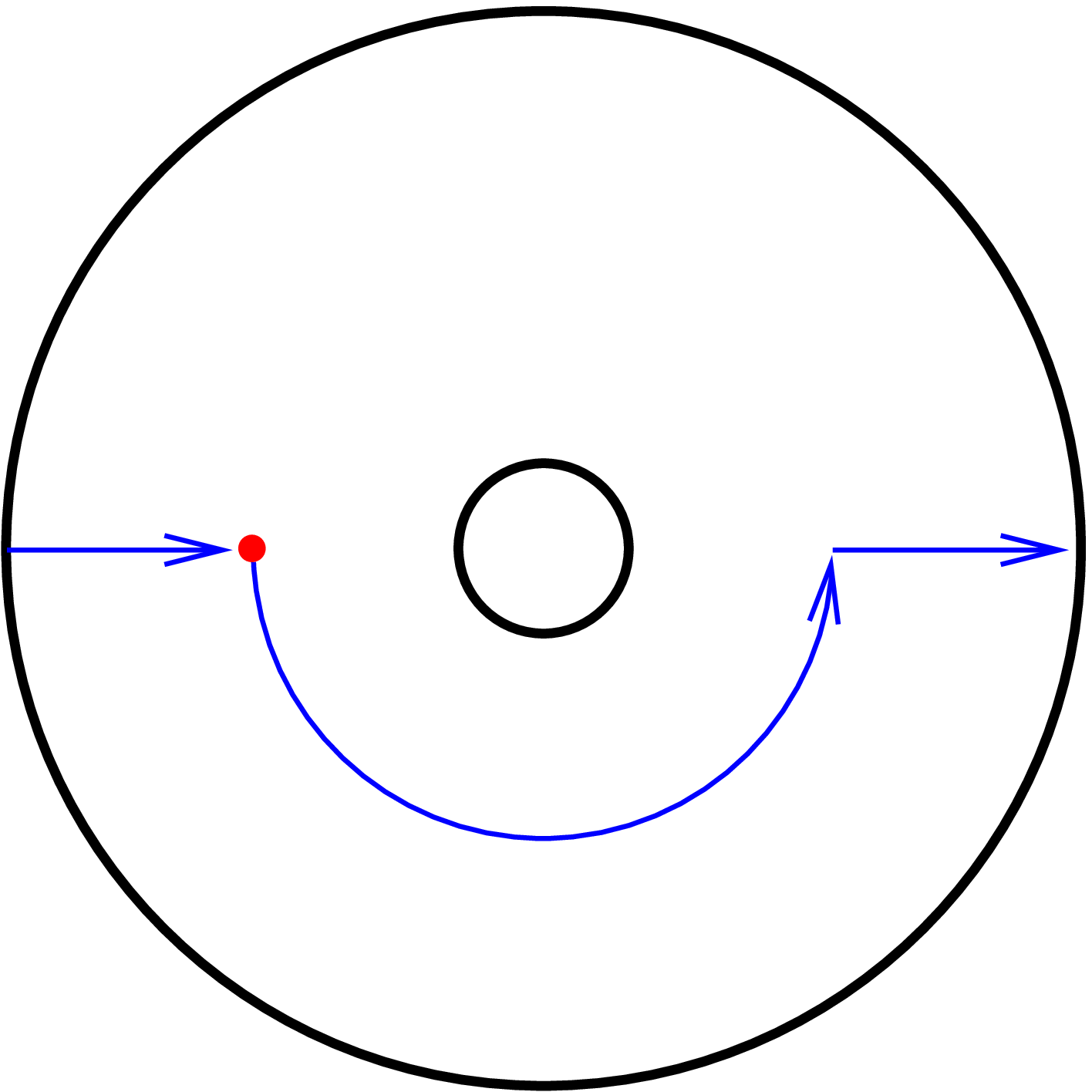,width=0.9\linewidth}
\end{minipage}
\put(-286,-40){\Large $a$}
\put(-72,-42){\Large $b$}
\put(-327,0){$p$}
\put(-102,2){$p$}
\caption{\label{fig:PiOneGen}%
Loops $a$ (left picture) and $b$ (right picture) in 
$\RP^3\uplus\RP^3$ that generate the fundamental group 
$\mathbb{Z}_2*\mathbb{Z}_2=\langle a,b:a^2=b^2=1\rangle$
based at point $p$. The product loop, $c:=ab$ (first $a$ then $b$), 
is homotopic to the loop $\ell$ in Fig.\,\ref{fig:TwoRP3}, that is, 
to a circle fiber.}
\end{figure}

The fundamental group is the free product of two $\mathbb{Z}_2$: 
\begin{equation}
\label{eq:PiOneTwoRP3-Pres}
\pi_1(\RP^3\uplus\RP^3)
\cong\mathbb{Z}_2*\mathbb{Z}_2
=\langle a,b:a^2=b^2=1\rangle\,.
\end{equation}
Two loops representing the generators $a$ and $b$ are shown in 
Fig.\,\ref{fig:PiOneGen}.
\noindent
Their product, $ab=:c$, is homotopic to a circle fiber. Replacing the 
generator $b$ by $ac$ (recall $a^2=1$), the presentation 
(\ref{eq:PiOneTwoRP3-Pres}) can now be written in terms of  
$a$ and $c$: 
\begin{equation}
\label{eq:PiOneTwoRP3-AltPres}
\pi_1(\RP^3\uplus\RP^3)
=\langle a,c:a^2=1,\, aca^{-1}=c^{-1}\rangle
\cong \mathbb{Z}\rtimes\mathbb{Z}_2\,.
\end{equation}
where $\mathbb{Z}_2$ (generated by $a$) acts as the automorphism 
$c\mapsto c^{-1}$ on the generator $c$ of $\mathbb{Z}$. This 
corresponds to the structure of $\RP^3\uplus\RP^3$ as $S^1$-fiber 
bundle over $\RP^2$ with $\pi_1(\mathrm{base})$ acting on 
$\pi_1(\mathrm{fibre})$. Algebraically, the normal subgroup 
$\mathbb{Z}$ generated by $c$ is just the subgroup of words in $a,b$ 
containing an even number of letters.  

The generators of mapping classes are the (unique) exchange, $\omega$, 
the slide $\mu_{12}$ of prime 2 through prime 1 (there is only 
one generator of $\pi_1$ for each prime and hence a unique generating 
slide through each prime), and the slide $\mu_{21}$ of prime 1 through 
prime 2. The relations between them are $\omega^2=\mu^2_{12}=\mu^2_{21}=1$
and $\omega\mu_{12}\omega^{-1}=\mu_{21}$. There are no other relations, 
as one may explicitly check using the Fouxe-Rabinovitch 
relations~\cite{McCulloughMiller:1986}. The particle group $G^P$ is 
just $\mathbb{Z}_2$, generated by $\omega$, and the slide subgroup 
$G^S$ is $\mathbb{Z}_2*\mathbb{Z}_2$, generated by $\mu_{12}$ and 
$\mu_{21}$. We have 
$\MCGF=G^S\rtimes G^P=(\mathbb{Z}_2*\mathbb{Z}_2)\rtimes\mathbb{Z}_2$,
where $\mathbb{Z}_2$ acts on $=\mathbb{Z}_2*\mathbb{Z}_2$ by permuting 
the factors. 

\begin{figure}[ht]
\begin{minipage}[c]{0.48\linewidth}
\centering\epsfig{figure=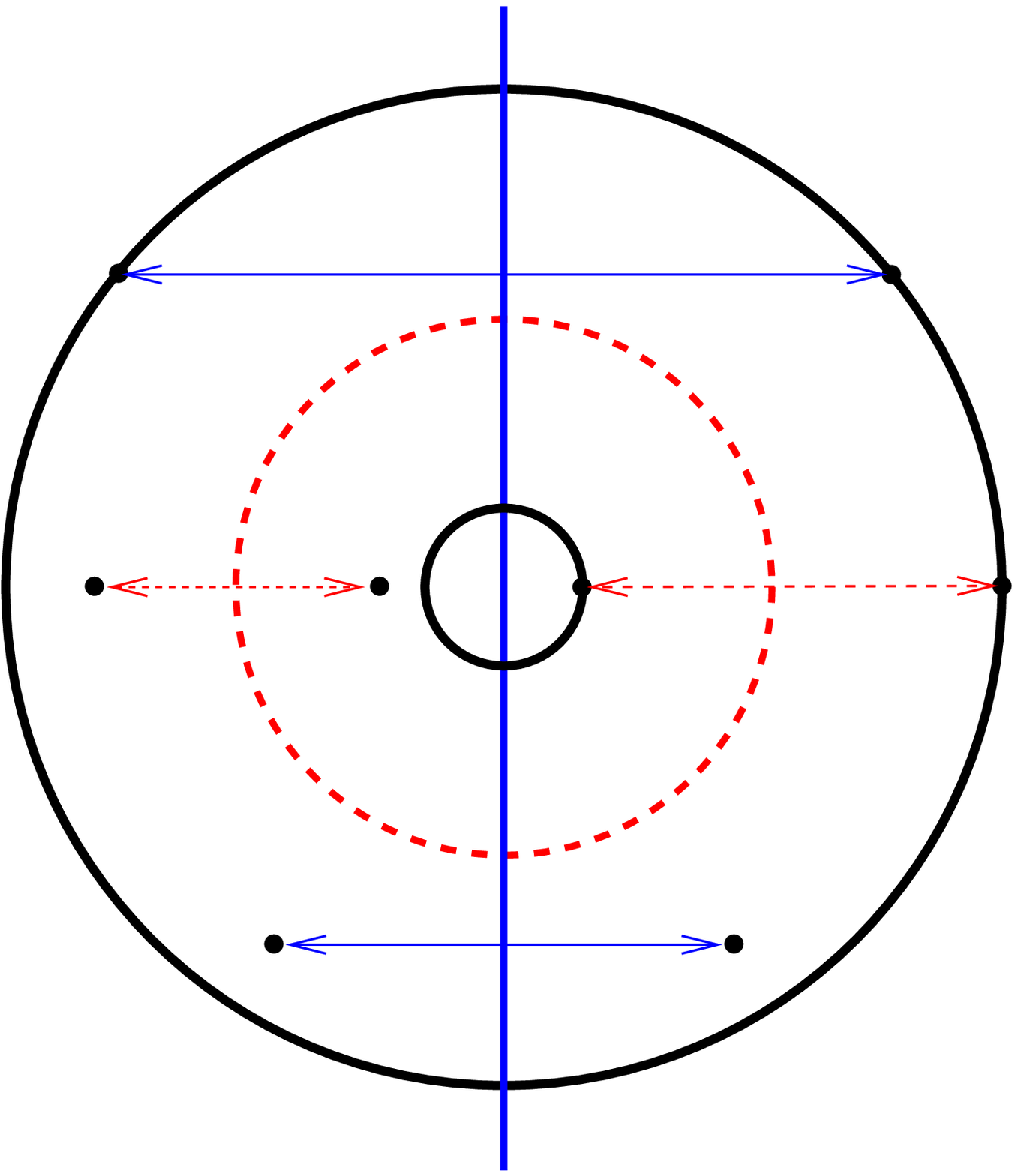,width=0.9\linewidth}
\end{minipage}
\hfill
\begin{minipage}[c]{0.48\linewidth}
\centering\epsfig{figure=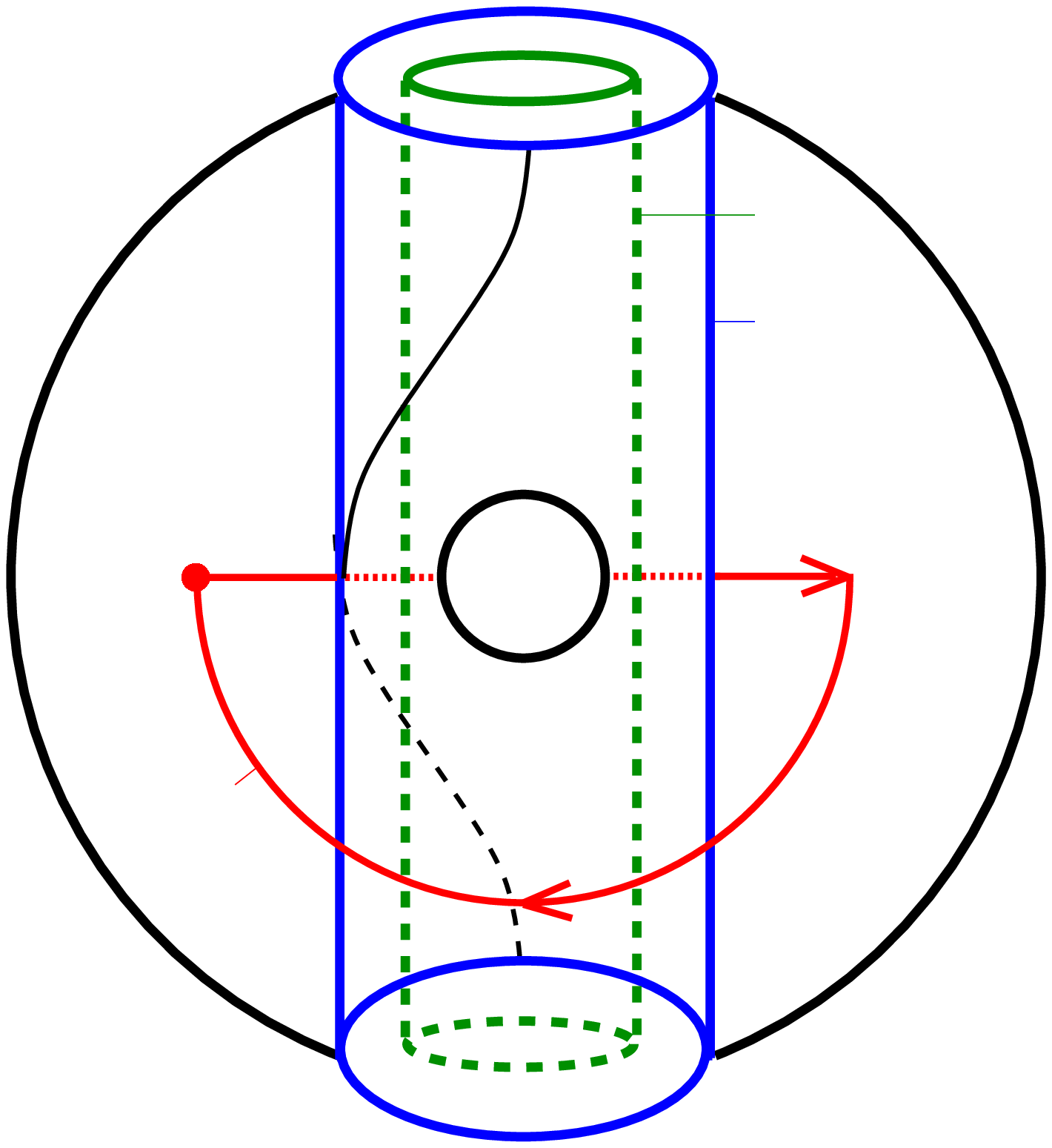,width=0.9\linewidth}
\end{minipage}
\put(-239,30){$S$}
\put(-272,95){$P$}
\put(-49,53){\footnotesize $T_1$}
\put(-49,38){\footnotesize $T_2$}
\put(-96,27){\tiny \begin{rotate}{60}$\theta=$ const.\end{rotate}}
\put(-142,0){\footnotesize $p$}
\put(-134,-32){\footnotesize $a$}
\caption{\label{fig:TwoRP3MCG}%
The left picture shows how the exchange generator $\omega$ can be 
represented by a combination of two reflections: the first reflection 
is at the connecting 2-sphere $S$ (dashed), whose action is 
exemplified by the dashed arrowed lines. This already exchanges 
the primes, though it is orientation reversing. In order to restore 
preservation of orientation we add a reflection at a vertical 
plane, $P$, whose action is exemplified by the solid arrowed lines. 
The right picture shows how the slide generator $\mu$ may be 
represented as the transformation of the form (\ref{eq:DefSlide}),
where the lines ($\theta=\mathrm{const}$) are now helical with 
a relative $\pi$-rotation between top and bottom, in order to be 
closed (due to the antipodal identification on $S_2$). We also 
draw the generator $a$ for the fundamental group on which the slide 
acts by conjugating it with $b$. The straight part pierces both 
tori whereas the curved part runs in front of them as seen from 
the observer.}
\end{figure}

Instead, we may get a presentation in terms of just two generators,
$\omega$ and $\mu:=\mu_{12}$, by dropping $\mu_{21}$ and the 
relation $\omega\mu_{12}\omega^{-1}=\mu_{21}$. This 
gives\footnote{This shows 
$(\mathbb{Z}_2*\mathbb{Z}_2)\rtimes\mathbb{Z}_2
\cong\mathbb{Z}_2*\mathbb{Z}_2$. This is no surprise. The normal 
subgroup isomorphic to $\mathbb{Z}_2*\mathbb{Z}_2$ of index 2 
is given by the set of words in the letters $\omega$, $\mu_{12}$, 
and $\mu_{21}$ containing an even number of the letter $\omega$.}  
\begin{equation}
\label{eq:TwoRP3MPG-Pres}
\mathcal{G}_{\mathrm{F}}(\RP^3\uplus\RP^3)
=\langle\omega,\mu:\omega^2=\mu^2=1\rangle
\cong \mathbb{Z}_2*\mathbb{Z}_2\cong\mathbb{Z}\rtimes\mathbb{Z}_2\,,
\end{equation}
where the last isomorphism follows as above 
(cf.(\ref{eq:PiOneTwoRP3-Pres},\ref{eq:PiOneTwoRP3-AltPres})). 
The normal $\mathbb{Z}$ is given by the subgroup of words in 
$\omega,\mu$ with an even number of letters.
A visualization of the generators $\omega$ and $\mu$ is attempted in 
Fig.\,\ref{fig:TwoRP3MCG}. It relies on the picture developed in 
 Fig.\,\ref{fig:TwoRP3}, where the region inside the 
sphere $S$ is identified with prime 1 and the region outside $S$
with prime 2. 

Following \cite{Aneziris-etal:1989a}, the set of inequivalent 
unitary irreducible representations (UIR's) can be determined 
directly as follows. First observe that under any UIR 
$\rho:(\omega,\mu)\mapsto(\hat\omega,\hat\mu)$ the algebra 
generated by $\hat\omega,\hat\mu$---we call it the `representor 
algebra'---contains $\hat\omega\hat\mu+\hat\mu\hat\omega$ in its 
center. To verify this, just observe that left multiplication of 
that element by $\hat\omega$ equals right multiplication by 
$\hat\omega$ (use $\hat\omega^2=1$), and likewise with $\hat\mu$. 
Hence, since $\rho$ is irreducible, this central element must be 
a multiple of the unit $\hat 1$ (by Schur's Lemma). This implies 
that the representor algebra is spanned by 
$\{\hat 1,\hat\omega,\hat\mu,\hat\omega\hat\mu\}$, i.e. 
it is four dimensional. Burnside's theorem (see e.g. \S\,10 in 
\cite{Weyl:GrQM}) then implies that $\rho$ is at most 2-dimensional. 
There are four obvious one-dimensional UIR's:   
\begin{subequations}
\label{eq:FourOneDimUIR}
\begin{alignat}{5}
\label{eq:FourOneDimUIR1}
& \rho_1:\ && \hat\omega &&\,=\,1\,,\quad &&\hat\mu &&\,=\,1\,,\\
\label{eq:FourOneDimUIR2}
& \rho_2:\ && \hat\omega &&\,=\,1\,,\quad &&\hat\mu &&\,=\,-1\,,\\
\label{eq:FourOneDimUIR3}
& \rho_3:\ && \hat\omega &&\,=\,-1\,,\quad &&\hat\mu &&\,=\,1\,,\\
\label{eq:FourOneDimUIR4}
& \rho_4:\ && \hat\omega &&\,=\,-1\,,\quad &&\hat\mu &&\,=\,-1\,.
\end{alignat}
\end{subequations}
The first two are bosonic the last two fermionic, either of them 
appears with any of the possible slide symmetries. 
The two dimensional representations are determined as follows: 
expand $\hat\omega$ and $\hat\mu$ in terms of 
$\{1,\sigma_1,\sigma_2,\sigma_3\}$, where the $\sigma_i$ are the 
standard Pauli matrices. That $\hat\omega$ and $\hat\mu$
each square to $\hat 1$ means that $\hat\omega=\vec x\cdot\vec\sigma$ and 
$\hat\mu=\vec y\cdot\vec\sigma$ with 
$\vec x\cdot\vec x=\vec y\cdot\vec y=1$. Using equivalences we may 
diagonalize $\hat\omega$ so that $\hat\omega=\sigma_3$. The remaining 
equivalences are then uniquely fixed by eliminating $y_2$ and ensuring 
$y_1>0$. Writing $y_1=\sin\tau$ and $y_3=\cos\tau$ we thus have : 
\begin{equation}
\rho_\tau(\omega)=\hat\omega=
\begin{pmatrix}
1&0\\
0&-1
\end{pmatrix}\,,\qquad
\rho_\tau(\mu)=\hat\mu=
\begin{pmatrix}
\cos\tau & \sin\tau\\
\sin\tau &-\cos\tau
\end{pmatrix}\,,
\quad \tau\in(0,\pi).
\end{equation}
This continuum of inequivalent 2-dimensional UIRs has interesting 
properties as regards the statistics types it describes. 
The angle $\tau$ mixes the bosonic and fermionic sector. This 
mixing is brought about by slides, which physically correspond to 
transformation where the two `lumps of topology' (geons) truly 
penetrate each other. Hence these geons have fixed statistics as 
long they do not come too close (at low energies), but cease to 
do so when deep scattering is involved. For further discussion 
we refer to \cite{Aneziris-etal:1989b} and especially 
\cite{SorkinSurya:1998}, where the case of more than two geons 
is considered. 

In other non-linear field theories in which to each 
kink-solution\footnote{We avoid the name soliton since we do 
not wish to imply that these objects are dynamically stable.}
there exists an anti-kink solution, general statements can be 
made regarding spin-statistics correlations~\cite{Sorkin:1988}. 
One would not expect such a relation to generalize to gravity 
without further specifications, since there is no such thing as 
an anti-geon. This follows immediately from the uniqueness of 
the prime decomposition. 

It has been verified that spinorial manifolds do in fact give 
rise to half-integer angular momentum states, e.g. in the 
kinematical Hilbert space of loop quantum 
gravity~\cite{ArnsdorfGarcia:1999} (see also \cite{Samuel:1993} 
for fractional spin in 2+1 dimensions). Hence a natural question 
is whether the existence or non-existence of spin-statistics 
violating states throws some light on the different schemes for the 
construction of states in quantum gravity. This has been looked 
at by Dowker \& Sorkin \cite{DowkerSorkin:1998}, who showed that 
the sum-over-histories approach excludes fermionic quantization of 
lens spaces (which are non-spinorial, as we have seen). Quite 
generally, they argue that topology changing amplitudes are 
necessary in order to avoid an embarrassing  abundance of sectors,
most of which are presumably unphysical. In \cite{DowkerSorkin:2000} 
the same authors discuss rules for assigning weights to individual 
histories and present three simple conditions of when to assign 
same weights that suffice to ensure the normal spin-statistics 
correlation. Unfortunately these rules seem too restrictive, in 
that they enforce the weights to exclusively come from 
\emph{abelian} representations of the mapping-class groups. This 
may altogether rule out spinorial states in the sum-over-histories 
approach, as it is often the case (any may be generally true) 
that the mapping-class generated by an overall $2\pi$-rotation is 
contained in the commutator subgroup and hence annihilated in any 
abelian representation.\footnote{This can be explicitly checked for 
spherical primes and the torus; see Section\,4 in \cite{Giulini:1995a}. 
Interestingly, the situation in 2+1 dimensions is quite different, 
as also discussed there.}

\section{Further  remarks on the general structure of $\MCGF$}
\label{sec:ResFin}
Generically, the group $\MCGF$ is non-abelian and of infinite
order; hence it will not be an easy task to understand its structure. 
We anticipated that the space of inequivalent UIRs label sectors 
in quantum gravity. However, this seems to only make sense if the 
group $\MCGF$ is of type\,I (see \cite{Mackey:1963,Mackey:UGR}), 
since only then can we  \emph{uniquely} decompose a unitary representation 
into irreducibles. It is known that a countable discrete group is of 
type\,I iff it contains an abelian normal subgroup of finite 
index~\cite{Thoma:1964}. This was indeed the case in our example 
above, where $\mathbb{Z}_2*\mathbb{Z}_2\cong\mathbb{Z}\rtimes\mathbb{Z}_2$,
hence $\mathbb{Z}$ is normal and of finite index. But, generically,  
being type\,I will be rather exceptional. 

Another important point is the following: suppose we argue (as e.g.
the authors of \cite{SorkinSurya:1998} do) that `internal' state spaces 
should be finite dimensional, in which case we would only be interested 
in finite dimensional representations of $\MCGF$. Are these sufficient 
to `make use' of each element of $\MCGF$? In other words: is any 
non-trivial element $\MCGF$ non-trivially represented in some finite 
dimensional representation? If not, the intersection of all kernels 
of finite dimensional representations would lead to a non-trivial 
normal subgroup and instead of $\MCGF$ we would only `see' its 
corresponding quotient. This question naturally leads to the general 
notion of residual finiteness. 
\begin{defn}
\label{def:ResFin}
A group $G$ is \emph{residually finite} iff for any non-trivial $g$ in $G$ 
there is a homomorphism $\rho_g$ into a finite group $F$ such that 
$\rho_g(g)$ is non-trivial in $F$. Equivalently, for each non-trivial 
$g$ in $G$ there exists a normal subgroup $N_g$ of $G$ (the kernel 
of $\rho_g$) of finite index such that $g\not\in N_g$.
\end{defn}

Residual finiteness is carried forward by various constructions. 
For example: 
\begin{itemize}
\item[1.)]
A subgroup of a residually finite group is residually finite
The proof is elementary and given in the appendix 
(Proposition\,\ref{prop:ResFinSub}).  
\item[2.)]
Let $G$ be the free product $G=G_1*\cdots *G_n$. Then $G$ is 
residually finite iff each $G_i$ is. For a proof see e.g. 
\cite{Gruenberg:1957}.
\item[3.)]
Let $G$ be finitely generated. If $G$ contains a residually finite 
subgroup of finite index then $G$ is itself residually finite.
The proof is given in the appendix 
(Proposition\,\ref{prop:ResFinFinIndSub}).
\item[4.)]
Let $G$ be finitely generated and residually finite. Then 
$\mathrm{Aut}(G)$ is residually finite. Again a proof is given in 
the appendix (Proposition\,\ref{prop:ResFinAut}).  
\end{itemize}

Note that 3.) implies that finite \emph{upward} extensions of residually 
finite groups (which are the normal subgroups) are residually finite. 
But, unfortunately, it is \emph{not} likewise true that finite 
\emph{downward} extensions (i.e. now the finite group is the normal 
subgroup) of residually finite groups are always residually finite
(see e.g.~\cite{Hewitt:1994}), not even if the extending group is 
as simple as $\mathbb{Z}_2$.\footnote{A simple (though not finitely 
generated) example is the central product of countably infinite copies 
of the 8 element dihedral group 
$D_8:=\langle a,b:a^4=b^2=(ab)^2=e\rangle
\cong\mathbb{Z}_4\rtimes\mathbb{Z}_2$, which can be thought of as 
the symmetry group of a square, where $a$ is the generator of the 
$\mathbb{Z}_4$ of rotations and $b$ is a reflection. Its center is 
isomorphic to $\mathbb{Z}_2$ and generated by $a^2$, the 
$\pi$-rotation of the square. In the infinite central product, 
where all centers of the infinite number of copies are identified 
to a single $\mathbb{Z}_2$, every normal subgroup of finite index 
contains the center. I thank Otto Kegel for pointing out this example.} 

We already mentioned above that if a group is residually finite 
the set of finite dimensional representations, considered as 
functions on the group, separate the group. Many other 
useful properties are implied by residual finiteness. For example, 
proper quotients of a residually finite group $G$ are never isomorphic
to $G$. In other words: any surjective homomorphism of $G$ onto $G$ 
is an isomorphism (such groups are called `Hopfian'). Most importantly, 
any residually finite group has a solvable word problem; see 
Proposition\,\ref{prop:ResFinSolWordProb} in the appendix.   

Large classes of groups share the property of residual finiteness. 
For example, all free groups are residually finite. Moreover, any group 
that has a faithful finite-dimensional representation in $GL(n,\mathbb{F})$, 
where $\mathbb{F}$ is a commutative field, i.e. any matrix group over a 
commutative field, is residually finite. On the other hand, it is also 
not difficult to define a group that is \emph{not} residually finite. 
A famous example is the group generated by two symbols $a,b$ and the 
single relation $a^{-1}b^2a=b^3$. Generally speaking, there are strong 
group-cohomological obstructions against residual 
finiteness~\cite{Hewitt:1995}. We refer to \cite{Magnus:1969} for 
an introductory survey and references on residual finiteness.  

Now we recall that the mapping class group $\MCGF$ is a finite downward 
extension of $h_{\mathrm{F}}\bigl(\mathrm{Aut}(\pi_1(\Sigma)\bigr)$, where 
$\pi_1(\Sigma)=\pi_1(P_1)*\cdots *\pi_1(P_N)$. Suppose each 
$\pi_1(P_i)$ ($1\leq i\leq N$) is residually finite, then so is 
$\pi_1(\Sigma)$ by 2.), $\mathrm{Aut}(\pi_1(\Sigma))$ by 4.), and 
$h_{\mathrm{F}}(\mathrm{Aut}(\pi_1(\Sigma))$ by 1.). So we must ask: 
are the fundamental groups of prime 3-manifolds residually finite?
They trivially are for spherical space forms and the handle. For the 
fundamental group of Haken manifolds residual finiteness has been 
shown in \cite{Hempel:1987}. Hence, by 3.), it is also true for 
3-manifolds which are virtually Haken, i.e. finitely covered by Haken 
manifolds. As already stated, it is conjectured that every irreducible 
3-manifold with infinite fundamental group is virtually Haken. If 
this were the case, this would show that all prime 3-manifolds, and 
hence all 3-manifolds\footnote{Recall that we restrict to compact and 
orientable(the latter being inessential here) manifolds.}, have 
residually finite fundamental group. 

Assuming the validity of the `virtually-Haken' conjecture (or else 
discarding those primes which violate it) we learn that 
$h_{\mathrm{F}}\bigl(\mathrm{Aut}(\pi_1(\Sigma)\bigr)$ is 
residually finite. This almost proves residually finiteness for
$\MCGF$ in case our primes also satisfy the 
homotopy-implies-isotopy property, since then $\MCGF$ is just 
a (downward) central $\mathbb{Z}^{n_s+m}_2$ extension of 
$h_{\mathrm{F}}\bigl(\mathrm{Aut}(\pi_1(\Sigma)\bigr)$, where 
$n_s$ is the number of spinorial primes and $m$ the number of 
handles. It can be shown that the handle twists (right picture in 
Fig.\,\ref{fig:DehnTwists}, which account for $m$ of the 
$\mathbb{Z}_2$s combine with 
$h_{\mathrm{F}}\bigl(\mathrm{Aut}(\pi_1(\Sigma)\bigr)$
into a semi-direct product (the extension splits with respect to 
$\mathbb{Z}^m_2$). Hence the result is residually finite by 3.). 
On the other hand, the remaining extension by the neck-twists 
(left picture in Fig.\,\ref{fig:DehnTwists}) certainly does 
not split and I do not know of a general argument that would 
also show residual finiteness in this case, though I would 
certainly expect it to hold. 

\section{Summary and outlook }
We have seen that mapping-class groups of 3-manifolds 
enter naturally in the discussion of quantum general relativity 
and, more generally, in any diffeomorphism invariant quantum 
theory. Besides being a fascinating mathematical topic in its 
own right, there are intriguing aspects concerning the physical 
implications of diffeomorphism invariance in the presence of 
non-trivial spatial topologies. Everything we have said is 
valid in any canonical approach to quantum gravity, may it be 
formulated in metric or loop variables. These approaches 
use 3-manifolds of fixed topology as fundamental entities 
out of which spaces of states and observables are to be built 
in a diffeomorphism (3-dimensional) equivariant way. Neither 
the spatial topology nor the spatial diffeomorphisms are 
replaced any anything discrete or quantum. Therefore the 
rich structures of mapping-class groups are carried along 
into the quantum framework. John Friedman and Rafael Sorkin 
were the first to encourage us to take this structure seriously 
from a physical perspective. Their work remind us on the old idea 
of Clifford, Riemann, Tait, and others, that otherwise empty space 
has enough structure to define localized matter-like properties: 
quasiparticles out of lumps of topology, an idea that was revived 
in the 1950s and 60s by John Wheeler and 
collaborators~\cite{Wheeler:Geometrodynamics}. 

The impact of diffeomorphism invariance is one of the central 
themes in all approaches to quantum gravity. The specific issue 
of mapping-class groups is clearly just a tiny aspect of it.
But this tiny aspect serves very well to give an idea of the range 
of possible physical implications, which is something that we 
need badly in a field that, so far, is almost completely 
driven by formal concepts. For example, the canonical approach 
differs in its wealth of sectors, deriving from 3-dimensional 
mapping classes, from the sum-over-histories approach.
In the latter, the mapping classes of the bounding 3-manifold do 
not give rise to extra sectors if they are annihilated after 
taking the quotient with respect to the normal closure of those 
diffeomorphisms that extend to the bulk; see e.g. 
\cite{HartleWitt:1988}\cite{GiuliniLouko:1992}. 
Other examples are the spin-statistics violating sectors, which 
have been shown to disappear in the sum-over-histories approach 
in specific cases~\cite{DowkerSorkin:1998}\cite{DowkerSorkin:2000}. 
However, whether the wealth of sectors provided by the canonical 
approach does indeed impose an `embarrassment of richness' from a 
physical point of view remains to be seen.  

\section*{Appendix: Elements of  residual finiteness}
For the readers convenience this appendix collects some of the 
easier proofs for the standard results on residual finiteness 
that where used in the main text. We leave out the proof for 
the result that a free product is residually finite iff each 
factor is, which is too involved to be presented here. 
The standard reference is~\cite{Gruenberg:1957}. 

In the following $H<G$ or $G>H$ indicates that $H$ is a subgroup 
of $G$ and $H\leftnormal G$ or $G\rightnormal H$ that $H$ is normal. 
The order of the group $G$ is denoted by $\vert G\vert$ and the 
index of $H$ in $G$ by $[G:H]$. The group identity will usually be 
written as $e$. The definition of residual finiteness was already 
given in Definition\,\ref{def:ResFin}, so that we can start 
with the first.
\begin{prop}
\label{prop:ResFinSub}
A subgroup of a residually finite group is again 
residually finite.
\end{prop}
\begin{proof}
Let $G$ be residually finite and $G'< G$. Hence, for 
$e\ne g'\in G'$ there exists a $K\leftnormal G$ of finite index 
such that $g'\not\in K$. Then $K':=K\cap G'$ is clearly a normal 
subgroup of $G'$ which does not contain $g'$. It is also of 
finite index in $G'$ since the cosets of $K'$ in $G'$ are given 
by the intersections of the cosets of $K$ in $G$ with $G'$. To 
see the latter, note that for $g'\in G'$ one has 
$g'K'= g'(K\cap G')=(g'K)\cap G'$, since $g'k\in G'$ iff 
$k\in G'$.
\end{proof}
\begin{lemma} 
\label{lemma:ResFinSubInt}
Let $G$ be a group and $H_i$, $i=1,\cdots,n$ a finite number 
of subgroups of finite indices. Then the intersection 
$H:=\bigcap_{i=1}^n H_i$ is itself of finite index.
\end{lemma}
\begin{proof} 
It suffices to prove this for two subgroups $H_1$ and $H_2$. 
We consider the left cosets of $H$, $H_1$ and $H_2$ and set 
$\vert G/H_i\vert=n_i$ for $i=1,2$. Elements $g,h\in G$ are
in the same $H$-coset iff $h^{-1}g\in H$, which is equivalent 
to $h^{-1}g\in H_i$ for $i=1,2$. Hence the $H$-cosets are 
given by the non-trivial intersections of the $H_1$-cosets 
with the $H_2$-cosets, of which there are at most $n_1\cdot n_s$.
\end{proof} 

\begin{lemma} 
\label{lemma:ResFinFinIndSub}
Let $G$ be finitely generated group, then the number of 
subgroups of a given finite index, say $n$, is finite.
\end{lemma}
\begin{proof} 
We essentially follow Chapter\,III in \cite{Baumslag:CombGrTh}. 
Let $H<G$ be of index $n$ and let $\rho(1),\cdots,\rho(n)$ a 
complete set of left-coset representatives, where without loss of 
generality we may choose $\rho(1)\in H$. The left cosets are then 
denoted by $\rho(i)H$ for $i=1,\cdots, n$. We now consider the 
($H$-dependent) homomorphism $\varphi:G\rightarrow S_n$, 
$g\mapsto\varphi_g$, of $G$ into the symmetric group of 
degree $n$, defined through 
$g\bigl(\rho(i)H\bigr)=:\rho\bigl(\varphi_g(i)\bigr)H$. 
Note that this just corresponds to the usual left action of $G$ on 
the left cosets, which we identified with the numbers 
$1,\cdots ,n$ via the choice of coset representatives. 
Since $gH=H\Leftrightarrow g\in H$ we have 
$\mathrm{stab}_{\varphi}(1):=\{g\in G:\varphi_g(1)=1\}=H$.
Now suppose $H'<G$ is also of index $n$. Repeating the construction 
above with left-coset representatives $g'(1),\cdots,g'(n)$ of $H'$, 
where $g'(1)\in H'$, we obtain another homomorphism 
$\varphi': G\rightarrow S_n$ with $\hbox{stab}_{\varphi'}(1)=H'$. 
Hence $H\not=H'\Rightarrow\varphi\not=\varphi'$. But since $G$ 
can be generated by a finite number of elements, say $m$, there 
are at most $(n!)^m$ different homomorphisms of $G$ into $S_n$, 
and hence at most $(n!)^m$ different subgroups of $G$ with index 
$n$. 
\end{proof}

For the following we recall that a subgroup $H<G$ is called 
\emph{characteristic} iff it is invariant under \emph{any} automorphism 
of $G$. Note that in case the group allows for non-trivial outer 
automorphisms this is a strictly stronger requirement than that of 
\emph{normality} which just requires invariance under inner automorphisms. 
We define 
\begin{subequations}
\label{eq:FinIndSubgr}
\begin{alignat}{2}
\label{eq:FinIndSubgr1}
&G_n     &&\,:=\, \bigcap\left\{H<G: [G:H]=n\right\}\,,\\
\label{eq:FinIndSubgr2}
& \bar{G}_n&&\,:=\, \bigcap\left\{H<G: [G:H]\leq n\right\}\,,
\end{alignat} 
\end{subequations}
i.e., the intersections of all subgroups of index $n$ or index 
$\leq n$ respectively. Lemma~\ref{lemma:ResFinFinIndSub}
ensures that there are only finitely many groups to intersect 
and Lemma~\ref{lemma:ResFinSubInt} guarantees that the intersection 
is again a group of finite index. Moreover, since an automorphism 
maps a subgroup of index $n$ to a subgroup of index $n$ it also 
leaves invariant the sets of subgroups of index $n$ or index 
$\leq n$ respectively. Hence we have
\begin{lemma}
\label{lemma:ResFinCharSubgroups}
$G_n$ and $\bar{G}_n$ are characteristic subgroups of finite 
index.
\end{lemma}

This can be used to give a convenient alternative characterization 
of residual finiteness for finitely generated groups:
\begin{prop}
\label{prop:ResFinAltDef}
Let $G$ be finitely generated. $G$ is residually finite 
$\Leftrightarrow$
\begin{equation}
\label{eq:ResFinAltDef}
\hat G:=\bigcap\left\{H<G: [G:H]<\infty\right\}=\{e\}
\end{equation}
\end{prop}
\begin{proof}
 ``$\Rightarrow$'': Residual finiteness implies that 
$g\not =e$ in $G$ is not contained in some normal subgroup of 
finite index. Hence it is not contained in $\hat G$. \\
``$\Leftarrow$'': For $g\not=e$ we have $g\not\in\hat G$ and hence 
$g\not\in G_n$ for some $n$. Since $G_n$ is in particular normal 
and of finite index, $G$ is residually finite.
\end{proof}

The following proposition is a conditional converse to 
Proposition\,\ref{prop:ResFinSub}: 
\begin{prop}
\label{prop:ResFinFinIndSub}
Let $G$ be finitely generated. If $G$ contains a residually finite 
subgroup $G'$ of finite index then $G$ is itself residually finite.
\end{prop}
\begin{proof} 
Let $[G:G']=k$, then $\bar{G}_n\subseteq G'$ for all
$n\geq k$. Hence
\begin{equation}
\label{eq:ResFinHat-G}
\hat G=\bigcap\left\{H<G : [G:H]<\infty\right\}
=\bigcap\left\{H<G': [G':H]<\infty\right\}=\{e\}\,,   
\end{equation}
where in the last step we applied Proposition\,\ref{prop:ResFinAltDef}
to $G'$. This is allowed if $G'$ is finitely generated, which is 
indeed the case since $G$ is finitely generated and $G'$ is of 
finite index. 
\end{proof}
Note that this proposition implies that finite upward extensions 
(cf. footnote\,\ref{foot:Extensions}) of residually finite groups 
are again residually finite. This is not true for finite downward 
extensions (see below). 

There are no analogs of Propositions\,\ref{prop:ResFinSub}
and \ref{prop:ResFinFinIndSub} for quotient groups. 
First, quotient groups of residually finite groups need not be 
residually finite. An example is provided by the free group 
$F_2$ on two generators, which is residually finite (as is any 
free group $F_n$, cf.~\cite{Gruenberg:1957}), but not its 
quotient group $\langle a,b: ab^2a^{-1}=b^3\rangle$ (see 
\cite{Magnus:1969}, p.\,307-308). Second, finite downward 
extensions of residually finite groups also need not be  
residually finite~\cite{Hewitt:1994}. 

We now turn to other important instances where residual 
finiteness is inherited.
\begin{prop} 
\label{prop:ResFinAut}
Let $G$ be a finitely generated and residually finite group, 
then $\mathrm{Aut}(G)$ is residually finite.
\end{prop}
\begin{proof} 
We follow Section\,6.5 of \cite{MagnusKarrassSolitar:CGT}.
Assuming $\hbox{Aut}(G)$ is non-trivial, let $\alpha$
be a non-trivial automorphism. Hence there exists a $g_*\in G$
such that $h:=g_*^{-1}\alpha(g_*)\not=e$. Residual finiteness 
of $G$ implies the existence of a $K\leftnormal G$ of 
finite index, say $n$, not containing $h$, so that $h\not\in G_n$ 
(cf.(\ref{eq:FinIndSubgr1})). On the other hand, since $G_n$ is 
characteristic, we have a natural homomorphism 
$\sigma:\hbox{Aut}(G)\rightarrow\hbox{Aut}(G/G_n)$, simply given 
by $\sigma(\varphi)(gG_n):=\varphi(g)G_n$, with kernel 
\begin{equation}
\label{eq:ResFinAut}
\kernel(\sigma)=\left\{\varphi\in\hbox{Aut}(G) : 
g^{-1}\varphi(g)\in G_n\,\forall g\in G\right\}\,.
\end{equation}
By Lemma\,\ref{lemma:ResFinCharSubgroups} $G/G_n$ is finite, 
and so $\hbox{Aut}(G/G_n)$ and  
$\image(\sigma)=\hbox{Aut}(G)/\kernel(\sigma)$ are finite, too. 
$h\not\in G_n$ now implies $\alpha\not\in\kernel(\sigma)$. 
Hence $\hbox{Aut}(G/G_n)$) is the sought for finite homomorphic 
image of $\hbox{Aut}(G)$ into which $\alpha$ maps non-trivially
via $\sigma$. 
\end{proof}

Finally we mention one of the most important consequences of
residual finiteness:  
\begin{prop}
\label{prop:ResFinSolWordProb}
Let $G$ be a finitely presented residually finite group, 
then it has a soluble word problem.
\end{prop}
\begin{proof} 
The idea is to construct two Turing machines, $T_1$ and $T_2$, 
which work as follows: Given a word $w$ on the finite set of 
generators, $T_1$ simply checks all consequences of the finite 
number of relations and stops if $w$ is transformed into $e$. 
So if $w$ indeed defines the neutral element in $G$, $T_1$ will 
eventually stop. In contrast, $T_2$ is now so constructed 
that it stops it $w$ is \emph{not} the neutral element. Using 
residual finiteness, this is possible as follows: $T_2$ writes 
down the image of $w$ under all homomorphisms of $G$ into all 
finite groups and stops if this image is not trivial. To do 
this it lists all finite groups and all homomorphisms into them 
in a `diagonal' (Cantor) fashion. To list all homomorphisms it 
lists all mappings from the finite set of generators of $G$ into 
that of the finite group, checking each time whether the relations 
are satisfied (i.e. whether the mapping defines a homomorphism).  
If $w$ does not define the neutral element, we know by residual 
finiteness that it will have a non-trivial image in some finite 
group and hence $T_2$ will stop after a finite number of steps. 
Now we run $T_1$ and $T_2$ simultaneously. After a finite number 
of steps either $T_1$ or $T_2$ will stops and we will know whether 
$w$ defines the  neutral element in $G$ or not.
\end{proof}


\newpage
\begin{thebibliography}{10}

\bibitem{Alvarez-GaumeGinsparg:1985}
Luis {Alvarez-Gaum\'e} and Paul Ginsparg.
\newblock The structure of gauge and gravitational anomalies.
\newblock {\em Annals of Physics}, 161:423--490, 1985.

\bibitem{Aneziris-etal:1989b}
Charilaos Aneziris et~al.
\newblock Aspects of spin and statistics in general covariant theories.
\newblock {\em International Journal of Modern Physics~A}, 14(20):5459--5510,
  1989.

\bibitem{Aneziris-etal:1989a}
Charilaos Aneziris et~al.
\newblock Statistics and general relativity.
\newblock {\em Modern Physics Letters~A}, 4(4):331--338, 1989.

\bibitem{ArnsdorfGarcia:1999}
Matthias Arnsdorf and Raquel~S. Garcia.
\newblock Existence of spinorial states in pure loop quantum gravity.
\newblock {\em Classical and Quantum Gravity}, 16:3405--3418, 1999.

\bibitem{Baumslag:CombGrTh}
Gilbert Baumslag.
\newblock {\em Topics in Combinatorial Group Theory}.
\newblock Lectures in Mathematics - ETH Z\"urich. Birkh\"auser, Basel, 1993.

\bibitem{BeigMurchadha:1987}
Robert Beig and Niall {\'O\,Murchadha}.
\newblock The {Poincar\'e} group as symmetry group of canonocal general
  relativity.
\newblock {\em Annals of Physics (NY)}, 174:463--498, 1987.

\bibitem{BroeckerJaenich:DiffTop}
Theodor Br\"ocker and Klaus J\"anich.
\newblock {\em {Einf\"uhrung in die Differentialtopologie}}.
\newblock Springer Verlag, Berlin, 1973.

\bibitem{CaboLouis-Martinez:1990}
Alejandro Cabo and Domingo Louis-Martinez.
\newblock On {Dirac's} conjecture for {Hamiltonian} systems with first-class
  constraints.
\newblock {\em Physical Review~D}, 42(8):2726--2736, 1990.

\bibitem{Atlas:FiniteGroups}
John~H. Conway et~al.
\newblock {\em {ATLAS} of Finite Groups}.
\newblock Oxford University Press, Oxford, 1985.

\bibitem{DeWittQTGI:1967}
Bryce~S. {DeWitt}.
\newblock Quantum theory of gravity. {I.\,The} canonical theory.
\newblock {\em Physical Review}, 160(5):1113--1148, 1967.

\bibitem{Dirac:LQM}
Paul Dirac.
\newblock {\em Lectures on Quantum Mechanics}.
\newblock Belfer Graduate School of Science, Monographs Series Number Two.
  Yeshiva University, New York, 1964.

\bibitem{DowkerSorkin:1998}
Fay Dowker and Rafael Sorkin.
\newblock A spin-statistics theorem for certain topological geons.
\newblock {\em Classical and Quantum Gravity}, 15:1153--1167, 1998.

\bibitem{DowkerSorkin:2000}
Fay Dowker and Rafael Sorkin.
\newblock Spin and statistics in quantum gravity.
\newblock In R.C. Hilborn and G.M. Tino, editors, {\em Spin-Statistics
  Connections and Commutation Relations: Experimental Tests and Theoretical
  Implications}, pages 205--218. American Institute of Physics, New York, 2000.

\bibitem{Fischer:1970}
Arthur~E. Fischer.
\newblock The theory of superspace.
\newblock In M.~Carmeli, S.I. Fickler, and L.~Witten, editors, {\em
  Relativity}, Proceedings of the Relativity Conference in the Midwest,
  Cincinnati, Ohio, June 2-6 1969, pages 303--357. Plenum Press, New York,
  1970.

\bibitem{Fischer:1986}
Arthur~E. Fischer.
\newblock Resolving the singularities in the space of {Riemannian} geometries.
\newblock {\em Journal of Mathematical Physics}, 27:718--738, 1986.

\bibitem{Fouxe-Rabinovitch:1940}
D.I. Fouxe-Rabinovitch.
\newblock {\"Uber die Automorphismengruppen der freien Produkte~I}.
\newblock {\em Matematicheskii Sbornik}, 8:265--276, 1940.

\bibitem{Fouxe-Rabinovitch:1941}
D.I. Fouxe-Rabinovitch.
\newblock {\"Uber die Automorphismengruppen der freien Produkte~II}.
\newblock {\em Matematicheskii Sbornik}, 9:297--318, 1941.

\bibitem{Freifeld:1968}
Charles Freifeld.
\newblock One-parameter subgroups do not fill a neighborhood of the identity in
  an infinite-dimensional lie (pseudo-) group.
\newblock In Cecile~M. DeWitt and John~A. Wheeler, editors, {\em Battelle
  Rencontres}, 1967 Lectures in Mathematics and Physics, pages 538--543. W.A.
  Benjamin, New York, 1968.

\bibitem{FriedmanSorkin:1980}
John Friedman and Rafael Sorkin.
\newblock Spin 1/2 from gravity.
\newblock {\em Physical Review Letters}, 44:1100--1103, 1980.

\bibitem{FriedmanWitt:1986}
John Friedman and Donald Witt.
\newblock Homotopy is not isotopy for homeomorphisms of 3-manifolds.
\newblock {\em Topology}, 25(1):35--44, 1986.

\bibitem{Gibbons:1986}
Gary~W. Gibbons.
\newblock The elliptic interpretation of black holes and quantum mechanics.
\newblock {\em Nuclear Physics}, B\,98:497--508, 1986.

\bibitem{Gilbert:1987}
Nick~D. Gilbert.
\newblock Presentations of the automorphims group of a free product.
\newblock {\em Proceedings of the London Mathematical Society}, 54:115--140,
  1987.

\bibitem{Giulini:1993}
Domenico Giulini.
\newblock On the possibility of spinorial quantization in the skyrme model.
\newblock {\em Modern Physics Letters A}, 8(20):1917--1924, 1993.

\bibitem{Giulini:1994a}
Domenico Giulini.
\newblock 3-manifolds for relativists.
\newblock {\em International Journal of Theoretical Physics}, 33:913--930,
  1994.

\bibitem{GiuliniLouko:1995}
Domenico Giulini.
\newblock Diffeomorphism invariant states in {Witten's} 2+1 quantum gravity on
  {$R\times T^2$}.
\newblock {\em Classical and Quantum Gravity}, 12:2735--2745, 1995.

\bibitem{Giulini:1995a}
Domenico Giulini.
\newblock On the configuration-space topology in general relativity.
\newblock {\em Helvetica Physica Acta}, 68:86--111, 1995.

\bibitem{Giulini:1995b}
Domenico Giulini.
\newblock Quantum mechanics on spaces with finite fundamental group.
\newblock {\em Helvetica Physica Acta}, 68:439--469, 1995.

\bibitem{Giulini:1995c}
Domenico Giulini.
\newblock What is the geometry of superspace?
\newblock {\em Physical Review\,D}, 51(10):5630--5635, 1995.

\bibitem{Giulini:1997a}
Domenico Giulini.
\newblock The group of large diffeomorphisms in general relativity.
\newblock {\em Banach Center Publications}, 39:303--315, 1997.

\bibitem{Giulini:1998}
Domenico Giulini.
\newblock On the construction of time-symmetric black hole initial data.
\newblock In F.W. Hehl, C.~Kiefer, and R.~Metzler, editors, {\em Black Holes:
  Theory and Observation}, volume 514 of {\em Lecture Notes in Physics}, pages
  224--243. Springer Verlag, Berlin, 1998.

\bibitem{Giulini:2003}
Domenico Giulini.
\newblock That strange procedure called quantisation.
\newblock In D.~Giulini, C.~Kiefer, and C.~L\"ammerzahl, editors, {\em Quantum
  Gravity: From Theory to Experimental Search}, volume 631 of {\em Lecture
  Notes in Physics}, pages 17--40. Springer Verlag, Berlin, 2003.

\bibitem{GiuliniLouko:1992}
Domenico Giulini and Jorma Louko.
\newblock Theta-sectors in spatially flat quantum cosmology.
\newblock {\em Physical Review D}, 46:4355--4364, 1992.

\bibitem{Goldhaber:1976}
Alfred~S. Goldhaber.
\newblock Connection of spin and statistics for charge-monopole composites.
\newblock {\em Physical Review Letters}, 36(19):1122--1125, 1976.

\bibitem{Gotay:1983}
Mark~J. Gotay.
\newblock On the validity of {Dirac's} conjecture regarding first-class
  secondary constraints.
\newblock {\em Journal of Physics~A: Mathematical and General}, 16:L141--L145,
  1983.

\bibitem{GotayNester:1984}
Mark~J. Gotay and James~M. Nester.
\newblock Apartheid in the {Dirac} theory of constraints.
\newblock {\em Journal of Physics~A: Mathematical and General}, 17:3063--3066,
  1984.

\bibitem{Gruenberg:1957}
Karl~W. {Gr\"unberg}.
\newblock Residual properties of infinite soluble groups.
\newblock {\em Proceedings of the London Mathematical Society}, 7:29--62, 1957.

\bibitem{HartleWitt:1988}
James Hartle and Donald Witt.
\newblock Gravitational {$\theta$}-states and the wave function of the
  universe.
\newblock {\em Physical Review~D}, 37(10):2833--2837, 1988.

\bibitem{HasenfratzHooft:1976}
Peter Hasenfratz and Gerard {'t\,Hooft}.
\newblock Fermion-boson puzzle in a gauge theory.
\newblock {\em Physical Review Letters}, 36(19):1119--1122, 1976.

\bibitem{Hatcher:1978}
Allan~E. Hatcher.
\newblock Linearization in 3-dimensional topology.
\newblock In Olli Lehto, editor, {\em Proceedings of the International Congress
  of Mathematicians}, volume~1, pages 463--468, Helsinki, 1978. American
  Mathematical Society (1980).

\bibitem{Hatcher:3-manifolds}
Allen~E. Hatcher.
\newblock Notes on basic 3-manifold topology.
\newblock Online available at
  www.math.cornell.edu/$\sim$hatcher/3M/3Mdownloads.html.

\bibitem{Hatcher:1976}
Allen~E. Hatcher.
\newblock Homeomorphisms of sufficiently large {$P^2$}--irreducible
  3-manifolds.
\newblock {\em Topology}, 15:343--347, 1976.

\bibitem{Hatcher:1983}
Allen~E. Hatcher.
\newblock A proof of the {Smale} conjecture, {Diff$(S^3)\simeq O(4)$}.
\newblock {\em Annals of Mathematics}, 117:553--607, 1983.

\bibitem{HawkingEllis:TLSSOS}
Stephen~W. Hawking and George~F.R. Ellis.
\newblock {\em The Large Scale Structure of Spacetime}.
\newblock Cambridge University Press, Cambridge, 1973.

\bibitem{Hempel:ThreeManifolds}
John Hempel.
\newblock {\em 3-Manifolds}.
\newblock Princeton University Press, Princeton, New Jersey, 1976.

\bibitem{Hempel:1987}
John Hempel.
\newblock Residual finiteness for 3-manifolds.
\newblock In S.M. Gersten and J.R. Stallings, editors, {\em Combinatorial Group
  Theory and Topology}, volume 111 of {\em Annals of Mathematics Studies},
  pages 379--396. Princeton University Press, Princeton, New Jersey, 1987.

\bibitem{Hendriks:1977}
Harrie Hendriks.
\newblock {Application de la th{\'e}orie d'obstruction en dimension~3}.
\newblock {\em M{\'e}moires de la Soci{\'e}t{\'e} Math{\'e}matique de France},
  53:81--196, 1977.
\newblock Online available at www.numdam.org.

\bibitem{HendriksMcCullough:1987}
Harrie Hendriks and Darryl McCullough.
\newblock On the diffeomorphism group of a reducible manifold.
\newblock {\em Topology and its Applications}, 26:25--31, 1987.

\bibitem{HenneauxTeitelboim:QGS}
Marc Henneaux and Claudio Teitelboim.
\newblock {\em Quantization of Gauge Systems}.
\newblock Princeton University Press, Princeton, New Jersey, 1992.

\bibitem{Hewitt:1994}
Paul~R. Hewitt.
\newblock Extensions of residually finite groups.
\newblock {\em Journal of Algebra}, 163(1):757--771, 1994.

\bibitem{Hewitt:1995}
Paul~R. Hewitt.
\newblock An army of cohomology against residual finiteness.
\newblock In C.M. Campbell, T.C. Hurley, E.F. Robertson, S.J. Tobin, and J.J.
  Ward, editors, {\em Groups `93}, volume 212 of {\em London Mathematical
  Society, Lecture Notes}, pages 305--313. Cambridge University Press,
  Cambridge, 1995.

\bibitem{Isham:1982}
Christopher~J. Isham.
\newblock {$Theta$}--states induced by the diffeomorphism group in canonically
  quantized gravity.
\newblock In J.J. Duff and C.J. Isham, editors, {\em Quantum Structure of Space
  and Time}, Proceedings of the Nuffield Workshop, August 3-21 1981, Imperial
  College London, pages 37--52. Cambridge University Press, London, 1982.

\bibitem{KalliongisMcCullough:1996}
John Kalliongis and Darryl McCullough.
\newblock Isotopies of 3-manifolds.
\newblock {\em Topology and its Applications}, 71(3):227--263, 1996.

\bibitem{Kiefer:QuantumGravity}
Claus Kiefer.
\newblock {\em Quantum Gravity}, volume 124 of {\em International Series of
  Monographs on Physics}.
\newblock Clarendon Press, Oxford, 2004.

\bibitem{Kneser:1929}
Hellmuth Kneser.
\newblock {Geschlossene Fl\"achen in dreidimensionalen Mannigfaltigkeiten}.
\newblock {\em {Jahresberichte der deutschen Mathematiker Vereinigung}},
  38:248--260, 1929.

\bibitem{Kruskal:1960}
Martin~D. Kruskal.
\newblock Maximal extension of schwarzschild metric.
\newblock {\em Physical Review}, 119(5):1743--1745, 1960.

\bibitem{Louko:2000}
Jorma Louko.
\newblock Single-exterior black holes.
\newblock In J.~Kowalski-Glikman, editor, {\em Towards Quantum Gravity}, volume
  541 of {\em Lecture Notes in Physics}, pages 188--202. Springer Verlag,
  Berlin, 2000.

\bibitem{Mackey:1963}
George~W. Mackey.
\newblock Infinite-dimensional group representations.
\newblock {\em Bulletin of the American Mathematical Society}, 69:628--686,
  1963.

\bibitem{Mackey:UGR}
George~W. Mackey.
\newblock {\em Unitary Group Representations in Physics, Probability, and
  Number Theory}.
\newblock Advanced Book Classics. Addison-Wesley Publishing Company, Redwood
  City, California, 1989.
\newblock Originally published in 1978 as part of the Mathematics Lecture Notes
  Series by the Benjamin/Cummings Publishing Company.

\bibitem{Magnus:1969}
Wilhelm Magnus.
\newblock Residually finite groups.
\newblock {\em Bulletin of the American Mathematical Society}, 75(2):305--316,
  1969.

\bibitem{MagnusKarrassSolitar:CGT}
Wilhelm Magnus, Abraham Karrass, and Donald Solitar.
\newblock {\em Combinatorial Group Theory: Presentations of Groups in Terms of
  Generators and Relations}, volume XIII of {\em Pure and Applied Mathematics}.
\newblock Interscience Publishers, John Wiley and Sons Inc., New York, 1966.
\newblock There is a Dover reprint edition of this book.

\bibitem{McCarty:1963}
G.S. McCarty.
\newblock Homeotopy groups.
\newblock {\em Transactions of the American Mathematical Society},
  106:293--303, 1963.

\bibitem{McCullough:1990}
Darryl McCullough.
\newblock Topological and algebraic automorphisms of 3-manifolds.
\newblock In Renzo Piccinini, editor, {\em Groups of Homotopy Equivalences and
  Related Topics}, volume 1425 of {\em Springer Lecture Notes in Mathematics},
  pages 102--113. Springer Verlag, Berlin, 1990.

\bibitem{McCulloughMiller:1986}
Darryl McCullough and Andy Miller.
\newblock Homeomorphisms of 3-manifolds with compressible boundary.
\newblock {\em Memoirs of the American Mathematical Society}, 61(344), 1986.

\bibitem{McDuff:1978}
Dusa McDuff.
\newblock The lattice of normal subgroups of the group of diffeomorphisms or
  homeomorphisms of an open manifold.
\newblock {\em Journal of the London Mathematical Society}, 18 (2.
  series):353--364, 1978.

\bibitem{Milnor:1957}
John~W. Milnor.
\newblock Groups which act on {$S^n$} without fixed points.
\newblock {\em American Journal of Mathematics}, 79(3):623--630, 1957.

\bibitem{Milnor:1962}
John~W. Milnor.
\newblock A unique decomposition theorem for 3-manifolds.
\newblock {\em American Journal of Mathematics}, 84(1):1--7, 1962.

\bibitem{Milnor:KTheory}
John~W. Milnor.
\newblock {\em Introduction to Algebraic K-Theory}, volume~72 of {\em Annals of
  Mathematics Studies}.
\newblock Princeton University Press, Princeton, New Jersey, 1971.

\bibitem{MisnerWheeler:1957}
Charles Misner and John~A. Wheeler.
\newblock Classical physics as geometry: Gravitation, electromagnetism,
  unquantized charge, and mass as properties of curved empty space.
\newblock {\em Annals of Physics (NY)}, 2:525--660, 1957.

\bibitem{NelsonAlvarez-Gaume:1985}
Philip Nelson and Luis {Alvarez-Gaum\'e}.
\newblock Hamiltonian interpretation of anomalies.
\newblock {\em Communications in Mathematical Physics}, 99(1):103--114, 1985.

\bibitem{Plotnick:1986}
Steven Plotnick.
\newblock Equivariant intersection forms, knots in {$S^4$} and rotations in
  2-spheres.
\newblock {\em Transactions of the American Mathematical Society},
  296(2):543--575, 1986.
\newblock Online available at www.jstor.org.

\bibitem{Reidemeister:1935}
Kurt Reidemeister.
\newblock {Homotopieringe und Linsenr\"aume}.
\newblock {\em {Abhandlungen des Mathematischen Seminars der Universit\"at
  Hamburg}}, 11:102--109, 1935.

\bibitem{Rindler:1965}
Wolfgang Rindler.
\newblock Elliptic {Kruskal-Schwarzschild} space.
\newblock {\em Physical Review Letters}, 15(26):1001--1002, 1965.

\bibitem{Samuel:1993}
Joseph Samuel.
\newblock Fractional spin from gravity.
\newblock {\em Physical Review Letters}, 71(2):215--218, 1993.

\bibitem{Seifert:1932}
Herbert Seifert.
\newblock {Topologie dreidimensionaler gefaserter R\"aume}.
\newblock {\em Acta Mathematica}, 60:147--288, 1933.

\bibitem{Skyrme:1971}
Tony Hilton~Royle Skyrme.
\newblock Kinks and the {Dirac} equation.
\newblock {\em Journal of Mathematical Physics}, 12(8):1735--1743, 1971.

\bibitem{Smale:1959}
Stephen Smale.
\newblock Diffeomorphisms of the 2-sphere.
\newblock {\em Proceedings of the American Mathematical Society},
  10(4):621--626, 1959.

\bibitem{Sorkin:1986}
Rafael Sorkin.
\newblock Introduction to topological geons.
\newblock In P.G. Bergmann and V.~De Sabbata, editors, {\em Topological
  Properties and Global Structure of Space-Time}, volume B138 of {\em NATO
  Advanced Study Institutes Series}, page 249. D. Reidel Publishing Company,
  Dordrecht-Holland, 1986.

\bibitem{Sorkin:1988}
Rafael Sorkin.
\newblock A general relation between kink-exchange and kink-rotation.
\newblock {\em Communications in Mathematical Physics}, 115:421--434, 1988.

\bibitem{Sorkin:1989}
Rafael Sorkin.
\newblock Classical topology and quantum phases: Quantum geons.
\newblock In S.~De Filippo, M.~Marinaro, G.~Marmo, and G.~Vilasi, editors, {\em
  Geometrical and Algebraic Aspects of Nonlinear Field Theory}, pages 201--218.
  Elsevier Science Publishers B.V., Amsterdam, 1989.

\bibitem{SorkinSurya:1998}
Rafael Sorkin and Sumati Surya.
\newblock An analysis of the representations of the mapping class group of a
  multi-geon three-manifold.
\newblock {\em International Journal of Modern Physics A}, 13(21):3749--3790,
  1998.

\bibitem{Thoma:1964}
Elmar Thoma.
\newblock {\"Uber unit\"are Darstellungen abz\"ahlbarer, diskreter Gruppen}.
\newblock {\em Mathematische Annalen}, 153(2):111--138, 1964.
\newblock Online available via www.digizeitschriften.de.

\bibitem{Thomas:EllipticStructures}
Charles~B. Thomas.
\newblock {\em Elliptic Structures on 3-Manifolds}, volume 104 of {\em London
  Mathematical Society Lecture Notes Series}.
\newblock Cambridge University Press, Cambridge, 1986.

\bibitem{Weyl:GrQM}
Hermann Weyl.
\newblock {\em Gruppentheorie und Quantenmechanik}.
\newblock Wissenschaftliche Buchgesellschaft, Darmstadt, 1981.

\bibitem{Wheeler:Geometrodynamics}
John~A. Wheeler.
\newblock {\em Geometrodynamics}.
\newblock Academic Press, New York, 1962.

\bibitem{Whitehead:1941}
John Henry~Constantine Whitehead.
\newblock On incidence matrices, nuclei and homotopy types.
\newblock {\em Annals of Mathematics}, 42(5):1197--1239, 1941.

\bibitem{Witt:1986b}
Donald Witt.
\newblock Symmetry groups of state vectors in canonical quantum gravity.
\newblock {\em Journal of Mathematical Physics}, 27(2):573--592, 1986.

\bibitem{Witt:1986a}
Donald Witt.
\newblock Vacuum space-times that admit no maximal slices.
\newblock {\em Physical Review Letters}, 57(12):1386--1389, 1986.

\bibitem{Wolf:SpacesConstCurv}
Joseph~A. Wolf.
\newblock {\em Spaces of Constant Curvature}.
\newblock MacGraw-Hill Book Company, New York, 1967.

\end{thebibliography}
\end{document}